\documentclass[sigplan,screen,10pt]{acmart}
\usepackage{hyperref}

\usepackage{verbatim}
\usepackage{xspace}
\usepackage{tikz}
\usepackage{listings}
\usepackage{graphicx}
\usepackage{tikz}
\usetikzlibrary{arrows}
\usetikzlibrary{shapes.geometric}
\usetikzlibrary{shapes.multipart}
\usepackage{pgfplots}
\usepackage{tabularx, booktabs}
\usepackage{url}
\usepackage{algorithm}
\usepackage{algpseudocode}
\usepackage{amsmath}
\usepackage{multirow}
\usepackage{makecell}
\usepackage{enumitem}
\usepackage[export]{adjustbox}

\usepackage{subcaption} 

\algblock{ParFor}{EndParFor}
\algnewcommand\algorithmicparfor{\textbf{par\ for}}
\algnewcommand\algorithmicpardo{\textbf{do}}
\algnewcommand\algorithmicendparfor{\textbf{end\ par\ for}}
\algrenewtext{ParFor}[1]{\algorithmicparfor\ #1\ \algorithmicpardo}
\algrenewtext{EndParFor}{\algorithmicendparfor}

\newcommand{\sysname}[0]{\textsc{NextDoor}\xspace}
\newcommand{\transit}{transit\xspace}
\newcommand{\transits}{transits\xspace}
\newcommand{\Transit}{Transit\xspace}

\newcommand{\source}{source\xspace}

\newcommand{\addition}[1]{{#1}} 

\lstdefinelanguage{NextDoorAPI}
    {keywords={long, int, char, for, if, else, then, Vertex, Edge, Sampler, bool, return, while, true, float, true, false, void},
    keywordstyle=\textbf,
    sensitive=true,
    morecomment=[l][\color{greencomments}]{///},
    morecomment=[l][\color{greencomments}]{//},
    morecomment=[s][\color{greencomments}]{{(*}{*)}},
    morestring=[b]",
    stringstyle=\color{redstrings},
    escapeinside={(*}{*)}, 
    escapechar=|, xleftmargin=1.0ex,numbersep=4pt,
    emph={srcInit, steps, next, distinct, sampleSize},
    emphstyle={\color{red}},
    mathescape=true, numbers=left, numberstyle=\footnotesize,
    firstnumber=1,
    basicstyle=\ttfamily
    }
\newcounter{groupcount}
\pgfplotsset{
    draw group line/.style n args={5}{
        after end axis/.append code={
            \setcounter{groupcount}{0}
            \pgfplotstableforeachcolumnelement{#1}\of\datatable\as\cell{%
                \def\temp{#2}
                \ifx\temp\cell
                    \ifnum\thegroupcount=0
                        \stepcounter{groupcount}
                        \pgfplotstablegetelem{\pgfplotstablerow}{[index]0}\of\datatable
                        \coordinate [yshift=#4] (startgroup) at (axis cs:\pgfplotsretval,0);
                    \else
                        \pgfplotstablegetelem{\pgfplotstablerow}{[index]0}\of\datatable
                        \coordinate [yshift=#4] (endgroup) at (axis cs:\pgfplotsretval,0);
                    \fi
                \else
                    \ifnum\thegroupcount=1
                        \setcounter{groupcount}{0}
                        \draw [
                            shorten >=-#5,
                            shorten <=-#5
                        ] (startgroup) -- node [anchor=north] {#3} (endgroup);
                    \fi
                \fi
            }
            \ifnum\thegroupcount=1
                        \setcounter{groupcount}{0}
                        \draw [
                            shorten >=-#5,
                            shorten <=-#5
                        ] (startgroup) -- node [anchor=north] {#3} (endgroup);
            \fi
        }
    }
}
\newcommand{\spara}[1]{\vspace*{0.05in}\noindent\textbf{#1}}

\begin{document}
\date{}
\title{Accelerating Graph Sampling for Graph Machine Learning using GPUs}

\author{Abhinav Jangda}
\affiliation{%
  \institution{University of Massachusetts Amherst}
  \country{United States}
}
\author{Sandeep Polisetty}
\affiliation{%
  \institution{University of Massachusetts Amherst}
  \country{United States}
}
\author{Arjun Guha}
\affiliation{%
  \institution{Northeastern University}
  \country{United States}
}
\author{Marco Serafini}
\affiliation{%
  \institution{University of Massachusetts Amherst}
  \country{United States}
}

\acmYear{2021}\copyrightyear{2021}
\setcopyright{acmcopyright}
\acmConference[EuroSys '21]{Sixteenth European Conference on Computer Systems}{April 26--28, 2021}{Online, United Kingdom}
\acmBooktitle{Sixteenth European Conference on Computer Systems (EuroSys '21), April 26--28, 2021, Online, United Kingdom}
\acmPrice{15.00}
\acmDOI{10.1145/3447786.3456244}
\acmISBN{978-1-4503-8334-9/21/04}

\begin{abstract}
Representation learning algorithms automatically learn the features of data.
Several representation learning algorithms for graph data, such as DeepWalk, node2vec, and GraphSAGE, \emph{sample} the graph to produce mini-batches that are suitable for training a DNN.
However, sampling time can be a significant fraction of training time, and existing 
systems do not efficiently parallelize sampling.

Sampling is an ``embarrassingly parallel'' problem and may appear to lend itself to GPU acceleration, but the irregularity of graphs makes it hard to use GPU resources effectively.
This paper presents \sysname, a system designed to effectively perform graph sampling on GPUs.
\sysname employs a new approach to graph sampling that we call {\transit-parallelism}, which allows load balancing and caching of edges.
\sysname provides end-users with a high-level abstraction for writing a variety of graph sampling algorithms.
We implement several graph sampling applications, and show that \sysname runs them orders of magnitude faster than existing systems.
\end{abstract}

\maketitle


\section{Introduction}
\emph{Representation learning} is a fundamental problem in machine learning.
Its goal is to learn features of data instead of hand-engineering them.
Representation learning on graph data involves mapping vertices (or subgraphs) to a $d$-dimensional vector known as an \emph{embedding}.
The embedding is then used as a feature vector for other downstream graph machine learning tasks.
Graph representation learning is a fundamental step in domains such as social network analysis, recommendations, epidemiology, and more.


Several algorithms for graph representation learning first \emph{sample} the input graph to obtain mini-batches and then \emph{train} a deep neural network (DNN) or a graph neural network (GNN) based on the samples. 
Moreover, different learning algorithms require different sampling mechanisms.
For example, DeepWalk~\cite{deepwalk} and node2vec~\cite{node2vec} use variants of random walks.
In contrast, GraphSAGE~\cite{graphsage}, which Pinterest uses for recommendation~\cite{pinterest}, samples the $k$-hop neighborhood of a vertex and uses their attributes to learn an embedding for each vertex.

Although several systems effectively leverage GPUs for the DNN training step, the same is not true for the sampling step.
Graph sampling takes a significant portion of total training time in real-world applications.
\addition{Table~\ref{tab:sampling-overhead} shows the impact of graph sampling in existing GNNs.
In each epoch, a GNN first samples the input graph to obtain mini-batches and then trains the DNN.
Graph sampling is an irregular computation that is typically performed using the CPU, whereas training is performed on the GPU.
In our experiments, we found that graph sampling can take up to 62\% of an epoch's time.\footnote{We use a 16-core Intel Xeon Silver CPU and an NVIDIA Tesla V100 GPU.}
This bottleneck is further exacerbated if the CPU is attached to multiple GPUs and cannot produce enough samples to saturate them.}
Hence, accelerating graph sampling is important to improve the end-to-end training time.



\begin{table}
  \small
  \addition{

  \begin{tabularx}{0.57\linewidth}{l|c|c}
    \toprule
   Input Graphs                 &  PPI     & Reddit \\\hline
  GraphSAGE~\cite{graphsage}    &  51\%   & 45\%   \\
  FastGCN~\cite{fastgcn}        &  26\%    & 52\%\\
  LADIES~\cite{ladies}          &  40\%    & 62\%\\
  ClusterGCN~\cite{clustergcn}  & 4.1\%    & 24\%\\
  GraphSAINT~\cite{graphsaint}  & 25\%     & 30\% \\
  MVS~\cite{mvs}                & 24\%     & 25\%\\
  \bottomrule
  \end{tabularx}
  }
  \caption{Fraction of time spent in graph sampling 
  in training.\vspace{-2em}}
  \label{tab:sampling-overhead}
\end{table}



Since samples are drawn independently, graph sampling is an ``embarrassingly parallel'' problem that seems ideal for exploiting the parallelism of GPUs.
However, for a GPU to provide peak performance, the algorithm must be carefully designed to ensure regular computation and memory accesses, which is challenging on irregular graphs.
Several systems have been designed for random walks~\cite{knightking}, graph mining~\cite{arabesque, asap, pangolin, automine}, and graph analytics~\cite{gunrock,simd-x,tigr,subway}.
These systems consider samples (or subgraphs) as the fundamental unit of parallelism: they grow all samples in parallel by looking up the neighbors of the vertices of each sample.
However, such an approach leads to two issues in these systems: (i) irregular memory accesses and divergent control flow because consecutive threads can access the neighbors of different vertices, and
(ii) lower parallelism because computation on all vertices in a sample is performed serially by the thread responsible for growing the sample.

In this paper, we present \sysname, the first system to perform efficient graph sampling on GPUs.
\sysname introduces \emph{\transit-parallelism}, which is a new approach to parallel graph sampling. In \transit-parallelism, the fundamental unit of parallelism is a \emph{\transit vertex}, which is a vertex whose neighbors may be added to one or more samples of the graph. In \transit-parallelism, each \transit vertex is assigned to a group of threads such that each thread adds one neighbor of the \transit vertex to one sample.
With this technique we obtain better GPU execution efficiency due to lower warp divergence, coalesced global memory accesses, and caching of the \transit vertex edges in low-latency shared memory.
Thus the irregular computation on the graph is changed to a regular computation.
\sysname effectively balances load across \transit vertices, by assigning them to \emph{different kernels} based on the number of samples associated with a transit vertex.
Each kernel uses a different scheduling and caching strategy to maximize the usage of execution resources and memory hierarchy.
\sysname{} has a high-level API that abstracts away the low-level details of implementing sampling on GPUs and 
enables ML domain experts to write efficient graph sampling algorithms with few lines of code.
\addition{\sysname{} achieves significant speedups over state-of-the-art systems for graph sampling and improves training time of existing GNN systems by up to 4.75$\times$.}
The contributions of this paper are:
\begin{itemize}
  \item A high-level API for building graph sampling algorithms with efficient execution on GPUs (Section~\ref{sec:api}).
  \item A new \transit-parallel paradigm to perform graph sampling on GPUs (Section~\ref{sec:paradigm}).
  \item \sysname, which leverages \transit-parallelism and adds techniques for load balancing and caching of a \transit's adjacency list (Section~\ref{sec:system}).
  \item Performance evaluation of \sysname{} against state-of-the-art systems: (i) a system of writing random walks KnightKing~\cite{knightking}, (ii) existing GNNs (GraphSAGE~\cite{graphsage}, \addition{FastGCN~\cite{fastgcn}, LADIES~\cite{ladies}, GraphSAINT~\cite{graphsaint}, MVS~\cite{mvs}, ClusterGCN~\cite{clustergcn})}, and (iii) two graph processing frameworks Gunrock~\cite{gunrock} and Tigr~\cite{tigr} (Section~\ref{sec:eval}).
\end{itemize}

\sysname and our experimental setup is available at \url{https://plasma-umass.org/nextdoor-eurosys21/}.


\section{Background and Motivation}
\label{sec:background}

\subsection{Representation Learning on Graphs}
\begin{figure}[t]
  \centering
 \includegraphics[scale=0.55]{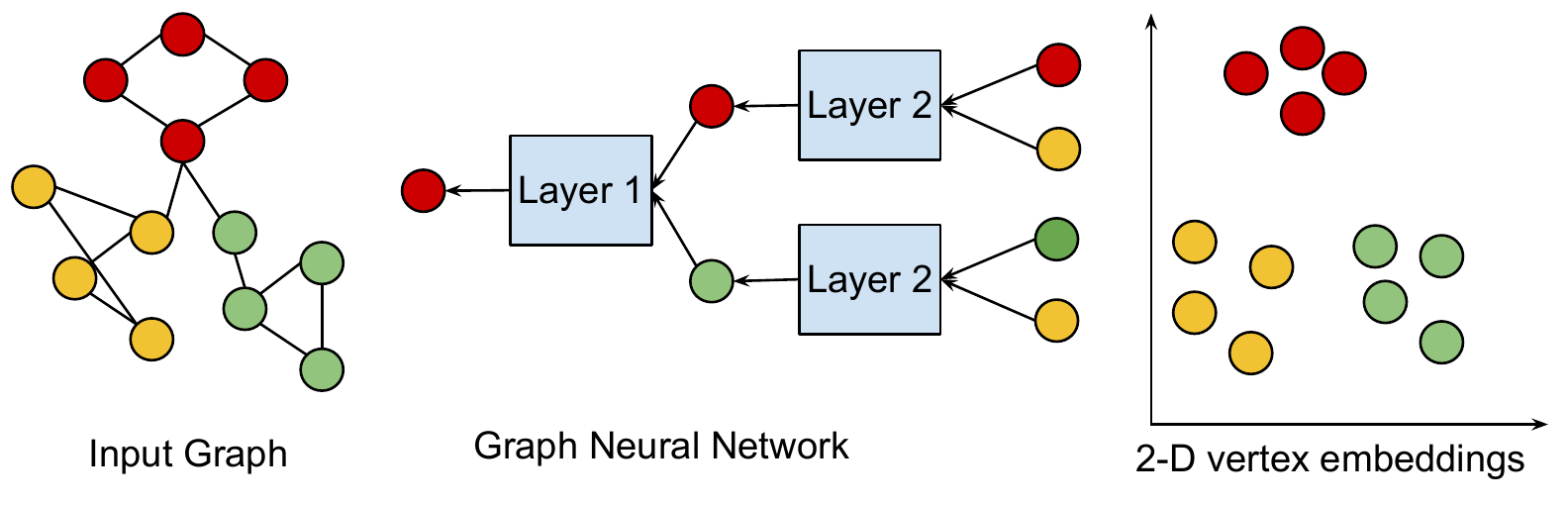}
 \caption{Representation learning on graphs. For each vertex of the input graph (on left), the Graph Neural Network (in middle) aggregates the information from $k$-hop neighbors back to the vertex.
 After training, each vertex is mapped to an embedding in a 2-dimensional space (on right).
}
 \label{fig:graph-embeddings}
\end{figure}

The goal of a graph representation learning algorithm is to map vertices (or subgraphs) to an embedding, which is a $d$-dimensional vector (Figure~\ref{fig:graph-embeddings}).

Early algorithms, such as DeepWalk~\cite{deepwalk} and node2vec~\cite{node2vec} employ \emph{shallow encodings}.
Given an input graph with $n$ vertices and a target $d$-dimensional Euclidean space, a shallow encoding is a $d \times n$ matrix where the $i^{th}$ column contains the embedding of vertex $v_i$.
These algorithms are \emph{transductive}: they take a static graph as input and produce embeddings only for the vertices in that graph.
They typically adapt the Skip-Gram approach~\cite{mikolov2013efficient} to graphs, performing random walks to obtain context and target vertices.

More recent algorithms, including GraphSAGE~\cite{graphsage}, are \emph{inductive}: they produce embeddings that
generalize to previously unseen vertices.
This property is particularly useful to build inference algorithms that work on dynamic, real-world graphs.
Inductive algorithms learn a \emph{deep encoding}, i.e., a function describing how to obtain a mapping, instead of the static map, and are also known as Graph Neural Networks (GNNs). 
Social networks like Pinterest uses GNNs to recommend newly added posts to users~\cite{pinterest}.

\addition{
During GNN training and inference, each vertex aggregates information from nodes in its $k$-hop neighborhood using a neural network.
Hops correspond to network layers, which are arranged as a tree.
Aggregation follows the tree from the $k$-hop vertices back to the root vertex (Figure~\ref{fig:graph-embeddings}).
Some GNN training systems perform whole-graph training, that is, they execute aggregation across the entire input graph without sampling~\cite{ma2019neugraph,jia2020improving}.
Most GNN algorithms, however, use mini-batching, where a mini-batch consists of a set of root vertices and a sample of their $k$-hop neighborhood (e.g.,~\cite{graphsage,fastgcn,ladies,graphsaint,mvs,clustergcn}).
The mini-batching approach based on sampling is easier to parallelize and scale to large graphs.
}

\subsection{Requirements for GPU Performance}

We now present an overview of modern GPU architectures and highlight characteristics of high-performance GPU code. These characteristics motivate the design of \sysname.

The fundamental unit of computation in a GPU is a \emph{thread}.
Threads are statically grouped into \emph{thread blocks} and assigned a unique ID within a block.
A GPU has multiple {\emph{streaming multiprocessors (SMs)}, each of which executes one or more thread blocks.
GPUs have several types of memory, two of which are relevant for this paper:
(1)~Each SM has a private memory, called \emph{shared memory}, which is only available to thread blocks assigned to that SM;
(2)~The GPU has \emph{global memory}, which is accessible to all SMs.
The access latency for global memory is significantly higher compared to shared memory and registers.

To run a thread block, an SM schedules a subset of threads from the thread block, known as a \emph{warp}.
A warp typically consists of 32 threads with consecutive thread IDs.
Moreover, GPUs employ a \emph{Single Instruction Multiple Threads (SIMT)} execution model:
all threads in a warp run the same instruction in lock-step.
One consequence of this execution model is that two threads cannot execute both sides of a branch concurrently.
Therefore, when the threads in a warp encounter a branch, the subset of threads that do not take the branch must wait for other threads to complete the branch.
This phenomenon is known as \emph{warp divergence} can lead to poor performance.
Thus, \emph{minimizing warp divergence} is key to achieving high-performance on GPUs.

It is also important to \emph{balance load across thread blocks}.
Suppose an SM is assigned to run thread blocks $A$ and $B$. Each thread block reserves a portion of SM resources, including registers and shared memory.
When $A$ is waiting, e.g., due to memory latency, the SM cannot switch execution to $B$ because the resources reserved by $A$ are unavailable to $B$.
(This behavior is very different from threads on a CPU, where context switches save registers to memory, thus all CPU registers are available to all threads.)
Hence, we need to \emph{balance resource usage across thread blocks} to concurrently execute the maximum number of thread blocks per SM.

Finally, a GPU program must explicitly choose to work with shared or global memory, and \emph{use shared memory when possible} to maximize performance.
In particular, when a thread waits on a memory access, it blocks all other threads in the same warp (another consequence of the SIMT execution model).
Therefore, the high latency of global memory access is particularly significant.
%
Fortunately, the GPU can provide high-bandwidth access to global memory by \emph{coalescing} several memory accesses from the same warp. This is only possible when \emph{concurrent memory accesses from threads in the same warp access consecutive memory segments}.


\section{An Abstraction for Graph Sampling}
\label{sec:graph-sampling-apps}

We introduce a general-purpose abstraction for graph sampling and use it to express common sampling algorithms.
The input to a graph sampling algorithm is a graph and an initial set of \emph{samples}, where each sample is a subset of vertices (and optionally edges) of the graph.
The algorithm iteratively grows each sample to include additional vertices in a series of \emph{steps}, and its output is the final set of expanded samples.
\addition{At each step, a sampling algorithm performs the following operations for each sample:}
\begin{enumerate}[leftmargin=*,topsep=0pt]
  \item \addition{Iteratively sample one vertex at a time and add it to the sample. This operation can access the neighborhood of some vertices, which we call \emph{\transit} vertices.}
  \item \addition{Determine the set of transit vertices for the next step.}
\end{enumerate}
\addition{
A graph sampling application can be expressed by providing user-defined functions that describe how to perform these operations.
Namely, the \emph{next} function describes how to sample one new vertex.
The \emph{samplingType} function describes the granularity at which we sample new vertices.
There are two types of sampling:}
\begin{enumerate}[leftmargin=*,topsep=0pt]
  \item  \addition{\emph{Individual \Transit Sampling:} the \emph{next} function is executed per-\transit a fixed number of times. It has access to the neighborhood of that \transit. }
  \item  \addition{\emph{Collective \Transit Sampling:} the \emph{next} function is executed per-sample a fixed number of times. It has access to the combined neighborhood of all \transit vertices. }
\end{enumerate}
\addition{
Finally, the \emph{stepTransits} function selects the vertices of the sample that will act as \transit vertices in the next step.}
Other user-defined parameters are the number of steps $k$, which could be set to $\infty$ if it varies from sample to sample, and the maximum number $m_i$ of \addition{new vertices sampled per \transit vertex (for individual \transit sampling) or per sample (for collective \transit sampling)} at step $i$.
We now show that by changing these user-defined functions and parameters, we can express 
a wide variety of sampling algorithms.

\begin{figure*}[t]
  \small
  \begin{subfigure}{0.13\textwidth}
   \includegraphics[scale=0.72]{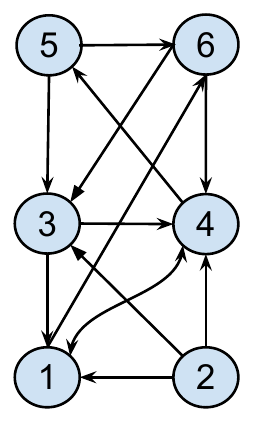}
   \caption{\addition{Graph}}
   \label{fig:graph-example}
\end{subfigure}
  \begin{subfigure}{0.44\textwidth}
    \centering
   \includegraphics[scale=0.72]{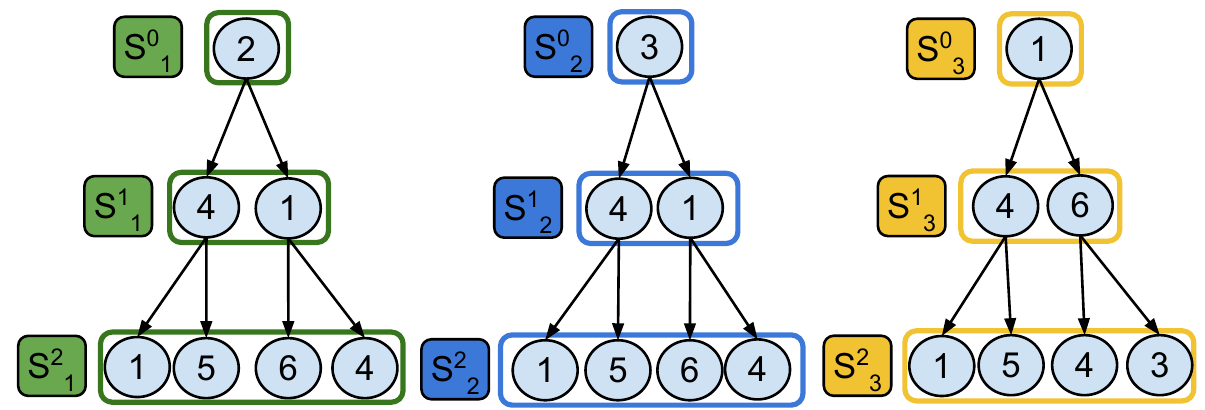}
   \caption{\addition{2-hop Neighborhood sampling}}
   \label{fig:k-hop-example}
  \end{subfigure}
  \hfill
  \begin{subfigure}{0.4\textwidth}
    \centering
   \includegraphics[scale=0.72]{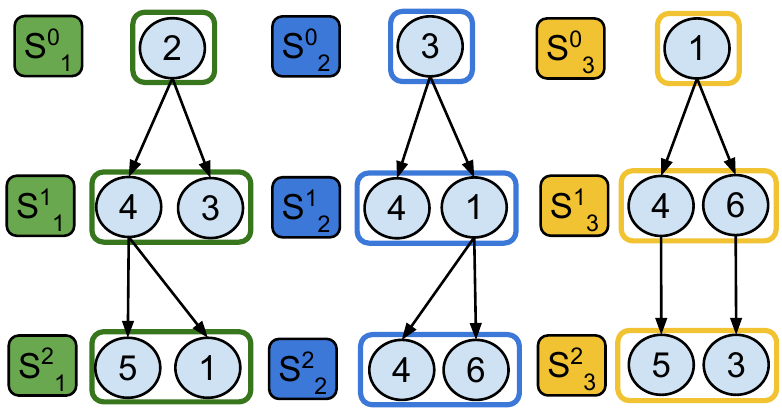} 
   \caption{\addition{Layer Sampling}}
   \label{fig:layer-sampling}
  \end{subfigure}
  \caption{\addition{Execution of a 2-hop Neighborhood sampling and Layer Sampling on graph in Figure~\ref{fig:graph-example} 
  for samples S$_1$, S$_2$, and S$_3$. For both applications the $next$ function uniformly samples
  two neighbors i.e., $m_1$ = 2 and $m_2$ = 2, and $\mathit{stepTransits}$ returns the vertices added in previous step as \transits.
  S$^{i}_j$ denotes the vertices obtained after step $i$ for sample S$_j$.
  Initially, $S_1$, $S_2$, and $S_3$ contains a single vertex.
  In the first step, the neighbors of \transit vertices, i.e., \textcircled{2}, \textcircled{3}, and 
  \textcircled{1} are added to the sample by both applications.
  While 2-hop Neighborhood choose two neighbors of each \transit vertex, Layer Sampling choose two neighbors from the set of all neighbors of all \transit vertices of a sample.
  In the second step, vertices sampled in first step becomes the \transit vertices. 
  Output of both sampling applications for each sample contains all vertices sampled at all steps.}}
\end{figure*}

\spara{Random walks}~\cite{deepwalk, node2vec,page-rank} A random walk starts from 
some initial set of root vertices.
In a static random walk, the probability of picking an edge is known beforehand, whereas in a dynamic random walk, the probability of picking an edge depends on properties of vertices that were previously visited on that walk.
DeepWalk~\cite{deepwalk} performs fixed-size biased static random walks, where the probability of following an edge is proportional to the edge weight.
Personalized Page Rank~\cite{personalized-page-rank}
performs a variable-size biased static random walk, where the probability of ending the random walk is defined by the user.
In contrast, node2vec~\cite{node2vec} is a dynamic random walk, which can be biased to stay closer to the starting vertex or to sample vertices that are further away.
\citet{knightking} provide a taxonomy of different types of random walks.

Our abstraction supports random walks as follows.
\addition{Random walks are individual \transit sampling applications because they sample a single neighbor of each \transit vertex of the sample at each step.}
Every element $m_i$ is $1$, since all random walks add at most one vertex at each step.
\addition{Since the \transit vertex is the previously sampled vertex, the $\mathit{stepTransits}$ function returns the previously sampled vertex in the sample.}
The root vertices are the initial samples, such that each sample is assigned one root vertex.
The number of steps $k$ describes the length of the fixed-size random walks in algorithms like DeepWalk and node2vec.
In these algorithms, $next$ always returns a vertex.
However, for applications that perform a variable-size walk, such as Personalized Page Rank, $k$ is set to $\infty$.
For termination, $next$ can decide to not add a new vertex to the sample.
The walk for a sample ends when the sample has no more new \transit vertices.

\spara{\addition{MultiDimensional Random Walks}}\addition{~\cite{multi-dimensional-rw} generalize regular random walks.
Each initial sample (walk) has a set of \emph{root} vertices, which are potential \transit vertices.
Each step extends the walk as follows.
First, a \transit vertex is selected from the set of root vertices. 
Then, a neighbor of the \transit vertex is added to the sample and replaces the \transit vertex in the set of roots. 
Multi-dimensional random walks can be represented in our abstraction as follows.
These walks perform individual \transit sampling with $k$ set to the length of walk, and each element $m_i$ to one.
The function $\mathit{stepTransits}$ returns a \transit vertex for a sample by randomly choosing one of the root vertices and 
then the $next$ function chooses a neighbor of the \transit vertex.
GraphSAINT~\cite{graphsaint} samples the graph using multi-dimensional random walks.}

\spara{$k$-hop Neighborhood Sampling}~\cite{graphsage} A $k$-hop neighborhood sampling algorithm employed in GraphSAGE~\cite{graphsage} adds one or more neighbors of a \transit vertex at each step.
Figure~\ref{fig:k-hop-example} shows the execution of a 2-hop Neighborhood sampler that samples two neighbors of each \transit at every step.
This is \addition{individual \transit sampling} and can be represented in our abstraction by setting $k = 2$, $m_1 = m_2 = 2$, \addition{having $\mathit{stepTransits}$ return all the vertices added in previous step as \transit}, and having \textit{next} uniformly choose neighbors of each \transit vertex and add them to the sample.

\spara{Layer Sampling}~\cite{layersampling} At each step $i$, layer sampling samples $m_i$ vertices from the set of neighbors of all \transit vertices of a sample, until the size of the sample reaches a maximum size ($M$) given by the user.
\addition{Unlike the sampling applications we have seen previously, layer sampling is a collective \transit sampling algorithm, where $m_i$ neighbors are chosen from the set of all neighbors of all \transit vertices of a sample.
The \transit vertices of a sample at a step are the vertices added in previous step to the sample.}
Since this sampling can run for arbitrary steps, we set $k$ to $\infty$ and the \textit{next} function chooses 
a neighbor from the set of all neighbors, if the size of the sample is less than $M$, otherwise \emph{next} does to not add a new vertex to the sample.
\addition{Figure~\ref{fig:layer-sampling} shows the execution of Layer Sampling that samples two edges from each \transit vertex.}
This can be represented in our abstraction by setting $k = 2$, $m_1 = m_2 =2$, \addition{having $\mathit{stepTransits}$ return the vertices added in previous step}, and 
having \textit{next} uniformly choose a neighbor from the combined neighborhood.

\section{Graph Sampling using \sysname}
\label{sec:api}

This section presents \sysname's API, which is based on our graph sampling abstraction
that we presented in the previous section. The API allows users to write a variety of graph sampling algorithms in just a few lines of code. Moreover, it allows users who are not experts in GPU programming to leverage modern GPU architectures.

\subsection{Programming API}

The inputs to \sysname are a graph, an initial set of samples, and several user-defined functions (Figure~\ref{fig:nextdoor-api}), which we detail below.
The output is an expanded set of samples.
If desired, \sysname can pick the initial set of samples automatically (e.g., select one random vertex per sample).

\addition{The user selects collective or individual \transit sampling using \texttt{samplingType}.
The \texttt{stepTransits} function returns the \transit vertices for a sample at a given step.
In individual \transit sampling, the number of \transit vertices for each sample at step $j$ are $\prod_{i=0}^{j} m_i$.
In collective \transit sampling, the number of \transit vertices for each sample are $m_{i-1}$.
This function takes three arguments: 1)~the step (\texttt{step}), 2)~the sample (\texttt{s}), and 3)~the index of \transit out of all \transits to return~(\texttt{transitIdx}).}
The user must also define a sampling function to use at each step of the computation (\texttt{next}). This function receives four arguments:
1)~the sample (\texttt{s}), 2)~\addition{the \source edge set to sample neighbors from (\texttt{srcEdges})}, 3)~\addition{\transit vertices
(\texttt{transits}) forming the \source edge set}, and 4)~the current step
(\texttt{step}).
\addition{If the sampling is individual \transit sampling then \texttt{transits} contains only a single \transit vertex and \texttt{srcEdges} contains the edges of this \transit vertex.
Otherwise, \texttt{transits} contains all \transit vertices of the sample and \texttt{srcEdges} contains edges of all \transit vertices}.
The result of \texttt{next} must be a vertex to add to $s$ (or a constant \texttt{NULL} that indicates not to add a neighbor).
The function \texttt{s.prevVertex(i, pos)} returns the vertex added at position \texttt{pos} of the last $i^{th}$ step, and
the function \texttt{s.prevEdges(i, pos)} returns the edges of that vertex.
This information is necessary for applications, such as node2vec.
The \texttt{steps} function defines the number of computational steps in the application ($k$). 
For applications that do not run for a fixed number of steps, such as Personalized Page Rank and Layer Sampling, they can return a special constant \texttt{INF} and the sampling process for a sample is stopped when no new \transit vertices are added to the sample.
The value returned by the \texttt{sampleSize} function determines
how many times the \texttt{next} function is invoked on each \addition{individual or collective} neighborhood for each sample at each step.
The \texttt{unique} function specifies if at a step the sample should contain only unique vertices.
The \texttt{Vertex} class has utility methods for computing the vertex degree, the maximum weight of all edges (\texttt{maxEdgeWeight}), and the prefix sum of all edges' weights.
Users can extend the class to include application-specific vertex attributes to be added to the samples.
\lstdefinelanguage{NextDoorAPI}
    {keywords={long, int, char, for, if, else, then, enum, SamplingType, Vertex, Edge, Sampler, Sample, Config, bool, return, while, true, float, true, false, void,Vector},
    keywordstyle=\textbf,
    sensitive=true,
    morecomment=[l][\color{green}]{///},
    morecomment=[l][\color{green}]{//},
    morecomment=[s][\color{green}]{{(*}{*)}},
    morestring=[b]",
    stringstyle=\color{redstrings},
    escapeinside={(*}{*)}, 
    escapechar=|, xleftmargin=1.0ex,numbersep=4pt,
    emph={samplingType, stepTransits, srcInit, steps, next, distinct, sampleSize, stepSize, previousSteps, stopSampling, unique},
    emphstyle={\color{red}},
    mathescape=true, numbers=left, numberstyle=\footnotesize,
    firstnumber=1,
    basicstyle=\ttfamily,
    keywordstyle=[2]{\color{purple}},
    otherkeywords = {previousEdges, previousVertex, maxEdgeWeight},
    morekeywords = [2]{previousEdges, previousVertex, maxEdgeWeight}
    }

\begin{figure}[t]
\small
\begin{lstlisting}[language=NextDoorAPI]
Vertex next(Sample s, Array<Vertex> transits,
            Array<Edge> srcEdges, int step);
int steps();
int sampleSize(int step);
bool unique(int step);
enum SamplingType {Individual, Collective};
SamplingType samplingType(); 
Vertex stepTransits(int step, Sample s, 
                    int transitIdx);
\end{lstlisting}
\caption{User defined functions required to implement a graph sampling application in \sysname{}}
\label{fig:nextdoor-api}
\end{figure}


\spara{Output format} 
\sysname supports two output formats based on the application. 
1)~\sysname can return an array of samples, such that each sample contains all \transit vertices sampled at all steps. This format is required by GNNs that use random walks and layer sampling.
2)~\sysname can return vertices sampled at each step in an individual array.
This format is required by GNNs that uses $k$-hop neighborhood sampling.
The arrays are stored in the GPU in both cases.

\subsection{Use Cases}

We now present the implementation of several graph sampling algorithms using
\sysname. 




\begin{figure*}[t]
  \footnotesize
  \begin{subfigure}{\columnwidth}
      \begin{lstlisting}[language=NextDoorAPI]
Vertex next(s, transits, srcEdges, step) {|\label{line:node2vec:next-begin}|
  Vertex t = s.prevVertex(2,0); |\label{line:get-last-stop}|
  Vector<Edge> tEdges = s.prevEdges(2,0); |\label{line:last-stop-edges}|
  float p = 2.0, q = 0.5; |\label{line:get-return-param}| |\label{line:get-in-out-param}|
  float maxW = transits[0].maxEdgeWeight();
  return rejection-smpl (transits[0], 
        srcEdges, maxW, t, tEdges, p, q);} |\label{line:node2vec:next-end}|
int steps() {return 100;} |\label{line:node2vec:steps}|
int sampleSize(step) {return 1;}|\label{line:node2vec:maxsize}|
bool unique(int step) {return false;}
SamplingType samplingType()
{return SamplingType::Individual;}
Vertex stepTransits(step, s, transitIdx)
{return s.prevVertex(1, transitIdx);}
      \end{lstlisting}
  \caption{node2vec random walk of length 100}
  \label{fig:node2vec-random-walk}
    \end{subfigure}
    \begin{subfigure}{\columnwidth}
      \begin{lstlisting}[language=NextDoorAPI]
Vertex next(s, transits, srcEdges, step) {|\label{line:sage:begin}|
  Vertex v = randInt(0, graph.vertices());
  for (auto trn : transits)
    if (trn.hasEdge(v)) 
      s.addEdge(step, trn, v);
  return v;} |\label{line:sage:end}|
int steps() {return 5;} |\label{line:sage:steps}|
int sampleSize(int step) {return 64;}|\label{line:sage:maxsize}|
bool unique(int step) {return false;}
SamplingType samplingType()
{return SamplingType::Collective;}
Vertex stepTransits(step, s, transitIdx)
{return s.prevVertex(1, transitIdx);}
              \end{lstlisting}
          \caption{\addition{Importance Sampling}}
          \label{fig:importance-sampling-uniform}
    \end{subfigure}
    \begin{subfigure}{\columnwidth}
      \begin{lstlisting}[language=NextDoorAPI]
Vertex next(s, transits, srcEdges, step) {|\label{line:node2vec:next-begin}|
  int idx = randInt(0, srcEdges.size());
  Vertex v = srcEdges[idx];
  s.roots.replace(transits[0], v);
  return v;} |\label{line:node2vec:next-end}|
int steps() {return 100;} |\label{line:node2vec:steps}|
int sampleSize(step) {return 1;}|\label{line:node2vec:maxsize}|
bool unique(int step) {return false;}
SamplingType samplingType()
{return SamplingType::Individual;}
Vertex stepTransits(step, s, transitIdx)
{return s.roots[randInt(0,s.numRoots())];}

      \end{lstlisting}
 \caption{\addition{Multi Dimensional Random Walk of length 100}}
  \label{fig:multi-dimension-random-walk}
    \end{subfigure}
    \begin{subfigure}{\columnwidth}
      \begin{lstlisting}[language=NextDoorAPI]
Vertex next(s, transits, srcEdges, step) {|\label{line:sage:begin}|
  int idx = randInt(0, srcEdges.size());
  return srcEdges[idx];} |\label{line:sage:end}|
int steps() {return 2;} |\label{line:sage:steps}|
int sampleSize(int step) {
  return (step == 0) ? 25 : 10;}|\label{line:sage:maxsize}|
bool unique(int step) {return false;}
SamplingType samplingType()
{return SamplingType::Individual;}
Vertex stepTransits(step, s, transitIdx)
{return s.prevVertex(1, transitIdx);}
              \end{lstlisting}
          \caption{GrapSAGE's 2-hop neighbors
          }
          \label{fig:2-hop-sage}
    \end{subfigure}
    \caption{Use Cases of \sysname}
\end{figure*}


\spara{node2vec} The node2vec algorithm is a second-order random walk. 
Let $v$ be \transit vertex and $t$ be the \transit vertex of the last step.
The probability of picking edge $(v,u)$ depends on hyperparameters $p$ and $q$, and is determined using three cases: (i) if $u = t$ then the probability is $p$, (ii) if $u \neq t$ and $u$ is a neighbor of $t$ then the probability is $1/q$, or (iii) if $u \neq t$ and $u$ is not a neighbor of $t$ then the probability is 1.
The next vertex is sampled using rejection sampling, which takes these parameters as input~\cite{knightking}.

Figure~\ref{fig:node2vec-random-walk} presents node2vec in
\sysname. 
\addition{The argument \texttt{transits} of \texttt{next} contains one \transit vertex, since random walk is an individual \transit sampling algorithm.}
Parameters $p$ and $q$ can be returned by a user-defined function or added as constants.
\texttt{next} performs rejection sampling (\texttt{rejection-smpl}), the details of which are discussed in~\cite{knightking}.
\addition{\texttt{stepTransits} returns the vertex added at previous step.}
\texttt{sampleSize} returns 1 because we add only one neighbor of \transit at each step.
\texttt{steps} returns the length of walk, i.e., 100.

\spara{$k$-hop neighbors}
Figure~\ref{fig:2-hop-sage}  implements GraphSAGE's 2-hop neighborhood sampler in \sysname.
\texttt{stepTransits} returns the vertices added at previous step as \transits.
\texttt{next}
retrieves the \transit vertex for this sample in the \texttt{transit} variable, and chooses a neighbor of the \transit vertex. 
Since this is a 2-hop sampling, \texttt{steps}
returns 2. GraphSAGE~\cite{graphsage} sets the number of neighbors as $m_1 = 10$ and $m_2 = 25$, as reflected in \texttt{sampleSize}.
\addition{MVS~\cite{mvs} is implemented in a similar way as it obtains 1-hop neighbors of all initial vertices in the sample.}

\spara{\addition{MultiDimensional Random Walk}} 
\addition{
Figure~\ref{fig:multi-dimension-random-walk} implements this sampling in \sysname.
At each step, \texttt{stepTransits} returns the \transit vertex of a sample by choosing one of the root vertices of the sample randomly and \texttt{next} samples a neighbor and replaces the root vertex with this neighbor.}

\spara{\addition{Importance Sampling}} 
\addition{In FastGCN~\cite{fastgcn} and LADIES~\cite{ladies} every sample includes an adjacency matrix that records the edges between vertices added in the previous step (the \transit vertices) and the current step.
At each step $i$, $m_i$ vertices are sampled from the graph according to a probability distribution and these vertices are added to the sample. 
Figure~\ref{fig:importance-sampling-uniform} implements importance sampling as a 
collective \transit sampling as returned by \texttt{samplingType}.
At each step 64 vertices are sampled.
\texttt{next} chooses a vertex of graph according to a probability distribution and adds an edge between the vertex and all \transits, if it exists.
The sampling criteria in~\cite{fastgcn,ladies} can be added in \texttt{next}. 
\texttt{stepTransits} returns the vertices sampled in the last step.}

\spara{\addition{Cluster Sampling}} \addition{ClusterGCN~\cite{clustergcn} sampling obtains an adjacency matrix between all vertices of one or more clusters. 
Figure~\ref{fig:importance-sampling-uniform} sketches the implementation where at each step an edge is recorded in a sample's adjacency matrix if the edge exists between any two \transits.}

\section{Paradigms for Graph Sampling on GPUs}
\label{sec:paradigm}

This section presents two paradigms for parallel graph sampling.
Existing systems for graph sampling~\cite{knightking,graphsage,fastgcn,clustergcn,mvs,ladies,graphsaint} use \emph{sample parallellism}.
We discuss its shortcomings on GPUs and propose an alternative \emph{transit parallel} paradigm.



\subsection{Sample-Parallelism}
\label{sec:sample-parallel}

Graph sampling is an ``embarrassingly parallel'' problem and the natural approach to
parallelization is to process each sample in parallel, which we call the
\emph{sample parallel paradigm}.
The approach is
analogous to subgraph parallel expansion in graph mining systems~\cite{arabesque,automine,pangolin}.
\addition{We now discuss how to apply this approach to \sysname{}'s applications, performing both individual and collective  \transit sampling.}


\spara{\addition{Individual \Transit Sampling}}
The naive way to implement sample parallelism \addition{for individual \transit sampling} on GPUs is to assign a thread to each sample but this limits the amount of parallelism to the number of samples.
\sysname's API enables a new, fine-grained approach to sample parallelism.
At each step $i$, we assign consecutive $m_i$ threads to a pair of sample and \transit. Each thread then calls the user-defined function (\texttt{next}) on its \transit.
Since consecutive threads add vertices to the same sample, this approach allows writes to \emph{global memory} to be coalesced.
The algorithm visits samples and their transits in parallel.
Figure~\ref{fig:sample-parallel-execution} shows an example of sample parallel execution for the second step of Figure~\ref{fig:k-hop-example}.
In this example, each sample is assigned to a thread block containing four threads.
Each thread samples one vertex for the assigned \transit
and writes this vertex to the output.

\begin{figure*}[t!]
  \begin{subfigure}[t]{0.49\textwidth}

 \includegraphics[scale=0.73]{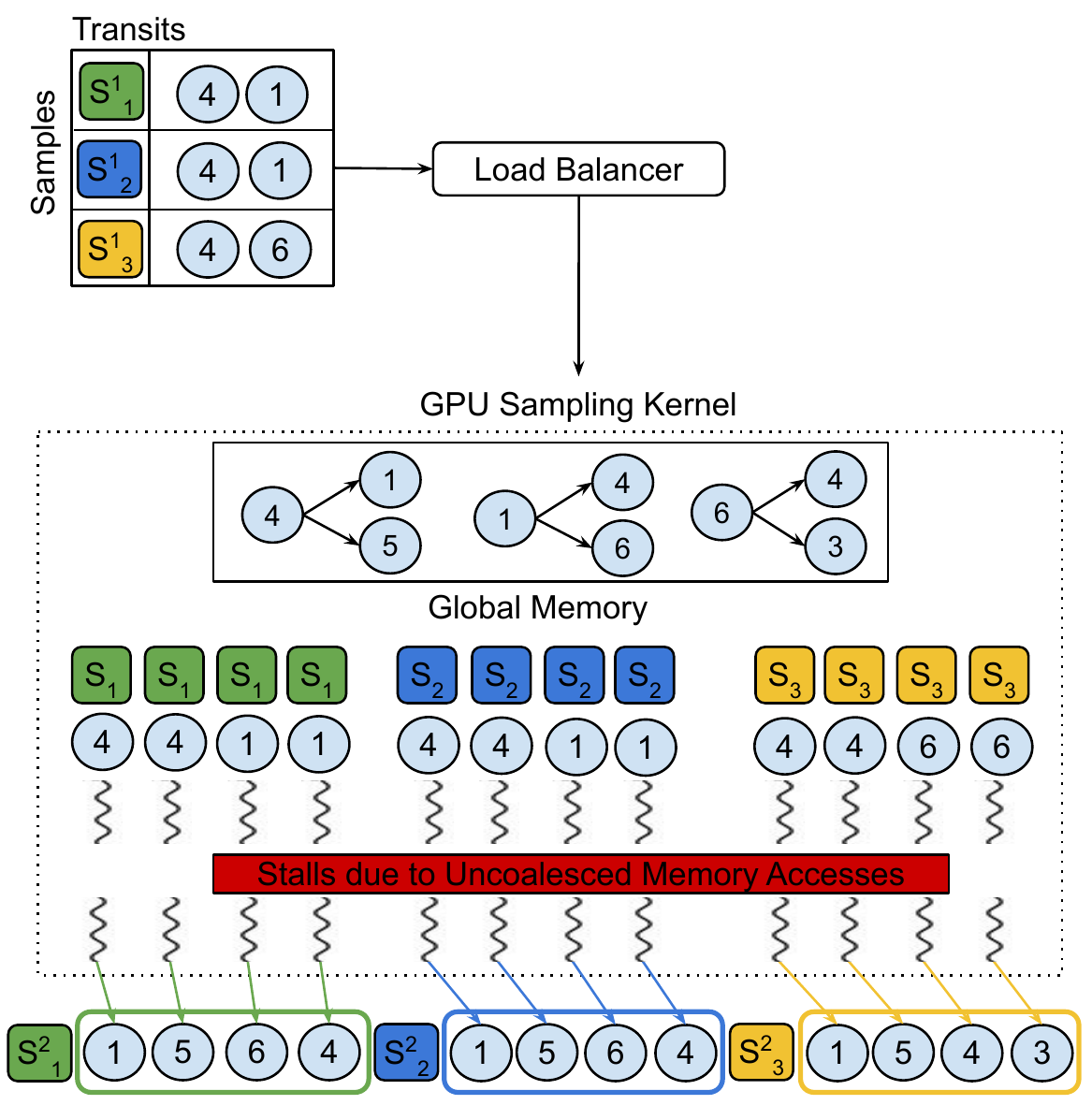}
 \caption{Sample Parallel execution of second step.
 Each pair of sample and \transit is assigned to consecutive threads.
 Each thread writes one vertex to the sample.
 Since contiguous threads can access different \transit vertices, we have divergent control flow and no locality.
 }
 \label{fig:sample-parallel-execution}
  \end{subfigure}
  \hfill
  \begin{subfigure}[t]{0.49\textwidth}
   \hspace{-7pt}
   \includegraphics[scale=0.73]{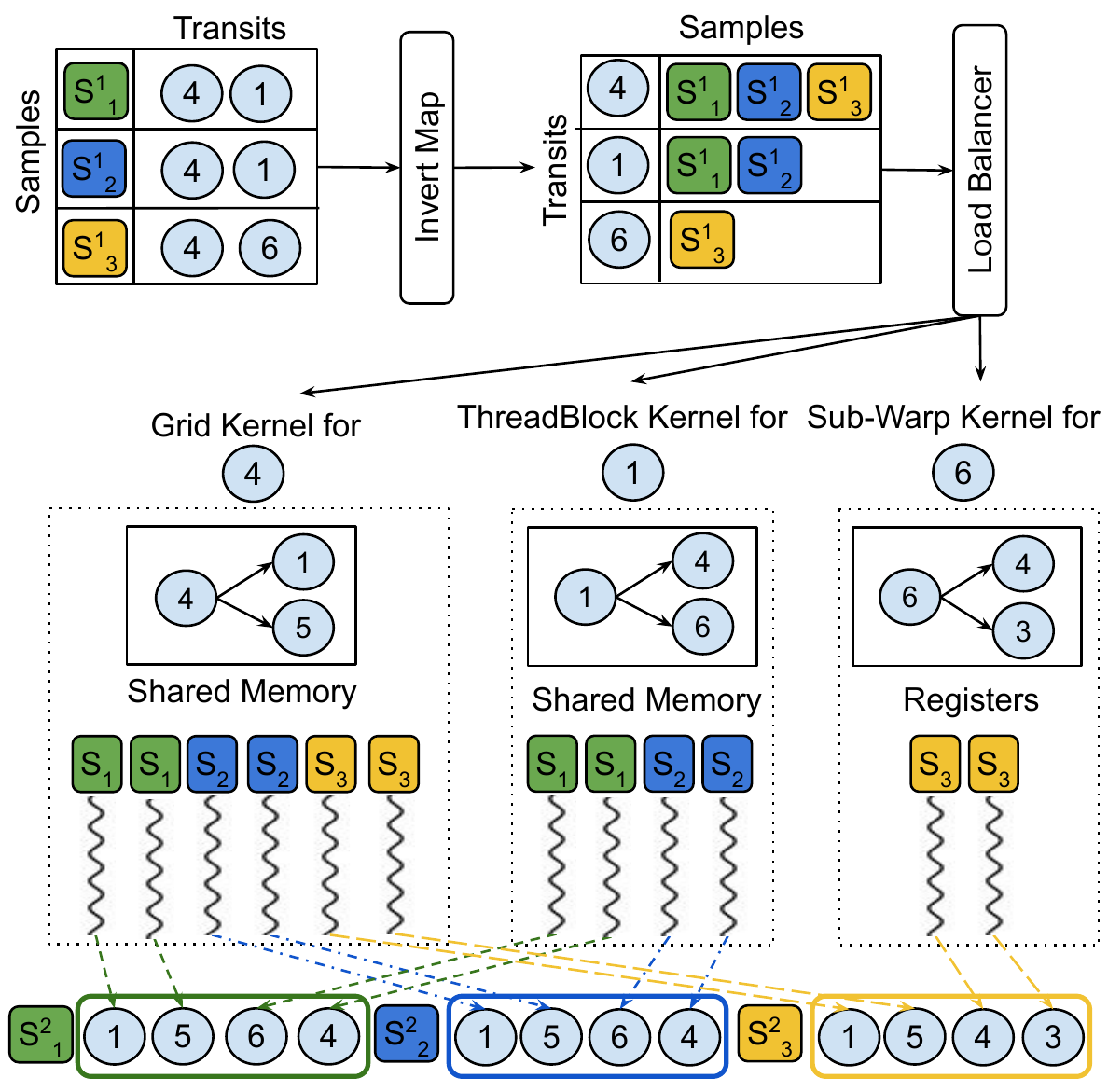}
   \caption{Transit Parallel execution of second step.
   Each \transit vertex is assigned to a group of threads: Grid, Thread Block, or Warp.
   Each thread writes a neighbor of the \transit vertex to the sample achieving non-divergent control flow and locality.
   }
   \label{fig:transit-parallel-execution-iter-2}
  \end{subfigure}
    \caption{Execution of second step for 2-hop neighbor sampling of Figure~\ref{fig:k-hop-example} using Sample Parallelism and Transit Parallelism.
    }
\end{figure*}

\spara{Collective \Transit Sampling} \addition{For collective \transit sampling, before calling the \texttt{next} function on the combined neighborhood of all \transits, we need to compute this neighborhood. 
In a sample parallel approach, the combined neighborhood is computed in the same way as the individual 	\transit sampling is performed.
Consecutive threads are assigned to each pair of sample and \transit, such that these threads copy the neighbors of one \transit to the combined neighborhood, which is stored in global memory. 
After computing the combined neighborhood, we assign each pair of sample and \transit  to $m_i$ consecutive threads and call \texttt{next} on this neighborhood.}

\spara{Limitations}
\addition{Despite the finer-grained approach enabled by the API, sample parallelism makes poor use of the GPU for the following reasons.}
1)~\addition{In an individual \transit sampling,} for each sample, the algorithm calls \texttt{next} on the neighbors of several \transit vertices in parallel.
However, if two threads in a warp are assigned to process two distinct \transit vertices with different numbers of neighbors, the thread processing the smaller set of neighbors may stall until the other thread completes. Thus the algorithm suffers from warp divergence.
\addition{Similarly, there is warp divergence when computing the combined neighborhood.}
2)~The algorithm also suffers from poor load balancing.
The amount of work done by \texttt{next} is likely to depend on the number of neighbors of the \transit vertex.
\addition{For example, while computing combined neighborhood in collective \transit sampling, different number of neighbors of each \transit vertex leads to load imbalance within a thread block.}
3)~The graph must be stored in global memory, so accessing neighbors of \transit vertex incurs high latency.
Moreover, threads in a block may access the neighbors
of different \transit vertices, which leads to no locality.
Hence, the GPU cannot coalesce reads or cache neighbors in shared memory.

For example, the execution in Figure~\ref{fig:sample-parallel-execution} suffers from both of the above issues.
Since all four threads do not process same \transit, there is divergent control flow and
the adjacency list must be stored in global memory, leading to lack of locality among all threads of a thread block.

\subsection{\Transit-Parallelism}
\label{sec:paradigm:transit-parallel}
To overcome the limitations of sample parallelism, we present the \emph{\transit parallel} paradigm.
\Transit parallelism groups all samples with the same \transit vertex and process all samples for one \transit vertex by assigning these samples to consecutive threads.
This approach exposes regularity in sampling.

At each step the \transit parallel paradigm works as follows.
Before running sampling on the GPU, we create a map of \transit vertices to their samples by grouping all samples associated with same \transit vertex.
We assign each \transit vertex to a group of threads, which may be organized as a grid, thread block, or warp.
\addition{In individual \transit sampling,} we assign each sample to consecutive threads in the group, and each thread calls \texttt{next} to add one neighbor of the \transit to its sample.
\addition{Similarly, in collective \transit sampling, we create the combined neighborhood of the \transits by assigning each sample to consecutive threads in the \transit group (grid, thread block, or warp), and consecutive threads in the group add neighbors of the \transit to the combined neighborhood of the sample.
Building the combined neighborhood in a sample-parallel manner takes a significant portion of execution time.
\sysname speeds up this step by using the \transit parallel approach.
In this case, instead of sampling new vertices from the neighborhood of each \transit of the sample using \texttt{next}, the system adds the entire neighborhood to the combined neighborhood.
Collective sampling applications then select new vertices from the combined neighborhood per sample.
}

Figure~\ref{fig:transit-parallel-execution-iter-2} shows the execution of the second step of 2-hop neighbor sampling of Figure~\ref{fig:graph-example} in \sysname using the \transit parallel paradigm.
Using load balancing (Section~\ref{sec:system}), \sysname assigns \transit \textcircled{4} to a grid, such that all thread blocks in that grid are assigned samples of \textcircled{4} and each thread adds one neighbor of \textcircled{4} to one sample of \textcircled{4}, i.e., either $S_1$, $S_2$, or $S_3$.
Similarly, it assigns vertex \textcircled{1} to a thread block and each thread adds one neighbor of \textcircled{1} to one of the samples associated with \textcircled{1}.

\spara{Advantages}
\addition{The \transit parallel paradigm 
has two advantages for both individual and collective \transit sampling}: 1)~contiguous threads perform a similar amount of work, because each thread calls the \text{next} function with same neighbors, and 2)~contiguous threads
accesses the neighbors of same \transit.
Both ensure non-divergent control flow, and locality of memory accesses.
This eliminates warp divergence and addresses load balancing \footnote{A badly-written user-defined function may have these issues, but \sysname{} avoids them in the core algorithm.}.
Moreover, since all threads work with same neighbors, we can cache the neighbors in shared memory to speed up later accesses.

For example, in the execution shown in Figure~\ref{fig:transit-parallel-execution-iter-2} each \transit is assigned to one group of threads and this group caches the neighbors of each \transit vertex in either shared memory or registers.
Furthermore, each thread calls the \textit{next} function on the same set of neighbors, which ensures non-divergent control flow among contiguous threads.

\section{Efficient Transit Parallelism on GPUs}
\label{sec:system}
\sysname implements transit parallelism on a GPU, with CPU-based coordination.
\addition{We first describe the techniques that allow \sysname{} to execute individual \transit sampling applications efficiently.
We then describe how \sysname{} uses the same techniques to execute collective \transit sampling applications.}

\subsection{Sampling in Individual \Transit Sampling}
\addition{In this section, we describe how \sysname{} executes individual \transit applications using \transit parallelism.}

\subsubsection{Leveraging Warp-Level Parallelism}
A GPU can \emph{coalesce} several global memory accesses together into one memory transaction only if threads in a warp access consecutive addresses.
The \transit parallel paradigm lends itself to a GPU implementation that supports coalescing \emph{reads} to global memory, by having consecutive threads read the same adjacency list (i.e., of the shared transit vertex).

However, coalescing \emph{writes} of new vertices to samples requires extra care.
A two-level \transit parallel approach maps different \transit vertices to thread blocks and different samples to threads.
This does not result in coalesced writes, since threads in the same warp add vertices to different samples.
Instead, \sysname uses three levels of parallelism: \transits to thread blocks, samples to warps, and a single execution of the \texttt{next} function to a thread.
Thus each thread writes one vertex to its sample and all threads in the warp issue one coalesced write to the same sample.
\addition{Figure~\ref{fig:transit-parallel-execution-iter-2} shows this mapping as follows. First, \transits \textcircled{4}, \textcircled{1}, and \textcircled{6} are mapped to a group of threads. Then, samples (S$_1$, S$_2$, S$_3$) are mapped to subwarps and each thread executes \texttt{next}.}

\spara{Sub-warps}
In an ideal scenario, there would be a one-to-one relationship between warps and samples, which would ensure that each thread in a warp writes to the same sample, using a single coalesced transaction to the global memory.
However, there is a fixed number of threads per warp (usually 32) and this number can sometimes be larger than the required number of executions of the \texttt{next} function.
Instead of letting threads be idle, \sysname shares a warp among several samples.
This yields some advantages.
Suppose we share a warp of 32 threads among 4 samples, each having 8 contiguous threads.
Then writes to the samples only generate 4 memory transactions rather than the 32 that we would obtain by assigning each thread to different sample.
This also does not lead to warp divergence because all threads in a warp sample neighbors of the same \transit vertex.

We use the term \emph{sub-warp} to refer to a set of contiguous threads of same warp assigned to same sample.
\sysname uses sub warps as a fundamental unit of resource scheduling.
All sub warps have the same size, which is determined using \texttt{sampleSize} function for the current step.
Threads of the same sub-warp share the information of their registers using \emph{warp shuffles}, and coordinate using \emph{syncwarp}.

\begin{table*}[t]
  \small
  \begin{tabularx}{53em}{lccccc}
      \toprule
        Kernel & Total neighbors to sample & Caching Strategy & Neighbor Access Strategy & \Transit Scheduling Strategy\\ \hline
        Grid & Greater Than 1024 & Shared Memory              & Memory Loads & One \transit to many thread blocks\\
        Thread Block & Between 32 and 1024 & Shared Memory    & Memory Loads & One \transit to one thread block\\
        Sub-Warp & Less than 32 & Registers                   & Warp Shuffles & One \transit to one sub-warp\\
      \bottomrule
  \end{tabularx}
  \caption{Types of kernels used to process transit vertices based on the number of neighbors to sample for the transit vertex.\vspace{-2em}}
  \label{tab:kernel-types}
\end{table*}

\subsubsection{Load Balancing}
\label{load-balancing}

In the \transit parallel paradigm, each \transit vertex is associated with a set of samples, which varies among \transit vertices and steps.
With three levels of parallelism, a \transit vertex requires as many threads in a step as the total number of neighbors that will be added to its samples.
Thus we obtain sub-optimal performance if we always assign a single thread block to each \transit vertex.
At the limit, the number of threads required by a \transit vertex may exceed the limit of number of threads in a block.
On the other hand, if a \transit vertex only requires a small number of threads, dedicating an entire thread block to the \transit wastes GPU resources.
To address this problem, \sysname uses three types of GPU kernel (see Table~\ref{tab:kernel-types}):
\begin{enumerate}[leftmargin=*,topsep=0in]

  \item The \emph{sub-warp} kernel processes several \transit vertices in a single warp. It is only applicable to \transit vertices that require fewer threads than the warp size (32).

  \item The \emph{thread block} kernel dedicates a thread block to a single \transit vertex. It is only applicable to \transit vertices that require more threads than in a warp, but less than the maximum thread block size (1,024).

  \item The \emph{grid} kernel processes a single \transit vertex in several thread blocks. It is only applicable to \transit vertex requires more than 1,024 threads.

\end{enumerate}

\spara{Scheduling}
To assign transits to kernels, \sysname{} creates a \textit{scheduling index} for each transit vertex.
Creating a scheduling index involves three stages.
First, \sysname creates a \transit-to-sample map based on the \transits obtained from the $\mathit{stepTransits}$ function (Figure~\ref{fig:transit-parallel-execution-iter-2}).
Then, \sysname partitions all \transit vertices into three sets based on the number of samples associated with each transit vertex using parallel scan operations.
Finally, the scheduling index of a transit vertex is set to the index of the transit vertex in its set.
After picking a kernel type for a \transit vertex, we assign each sample of the \transit vertex to a sub-warp in the kernel based on the thread index.

\spara{Caching} \sysname uses different caching strategies for different kernels to minimize memory access costs.
When sampling neighbors of \transit vertices in the \emph{grid} and \emph{thread block} kernels, the thread blocks for these kernels load the neighbors of \transit vertices into shared memory.
However, when the neighbors do not fit in shared memory, \sysname transparently
loads neighbors from global memory.
For \transit vertices assigned to a \emph{sub-warp}, \sysname{} utilizes both shared memory and thread local registers to store neighbors. 
In this case, \sysname{} transparently manages accesses to the neighbor list using \emph{warp shuffle} instructions that allows consecutive threads to read neighbors from each others' registers.
In summary, \sysname{} uses the fastest caching mechanisms available for each kernel.

\subsection{\Transit-Parallel Collective \Transit Sampling}
\addition{
Collective \transit sampling applications require computing the combined neighborhood of all the \transits of each sample.
This is a potential performance bottleneck, so \sysname{} uses transit parallelism to speed up the process.
It constructs the combined neighborhood as if it were an individual \transit sampling application that runs for only one step.
Instead of sampling new vertices from the neighborhood of one \transit, \sysname adds all the vertices in the neighborhood to the combined neighborhood of the sample.
After building a single combined \transit neighborhood per sample, one could in principle detect which samples have the same combined neighborhoods and expand all these samples in a transit-parallel manner.
The likelihood of two combined neighborhood being equal, however, is generally low, and detecting which samples have the same combined neighborhood is expensive.
Therefore, \sysname adds new vertices to the sample using a sample-parallel approach.}

\subsection{Unique Neighbors}
Certain applications, require all sampled neighbors from all \transit vertices to be unique.
After sampling at each step \sysname removes duplicated sampled vertices by first sorting them with a parallel radix sort, and then getting distinct vertices using parallel scan.
If sampled neighbors fit in the shared memory then \sysname performs this computation by assigning one sample to one thread block, otherwise one kernel is called for each sample.
After this process if the sample size is less than the \texttt{stepSize}, then \sysname performs sampling using a sample-parallel approach instead of the \transit-parallel one.

\subsection{Graph Sampling using Multiple GPUs}
\sysname utilizes the embarrassingly parallel nature of graph sampling to parallelize graph sampling over multiple GPUs in the following way.
First \sysname distributes samples equally among all GPUs.
Then, \sysname performs load balancing and scheduling and calls the sampling kernels on each GPU independently.
Finally, it collects the output from all GPUs.

\subsection{Integration in GNNs using Python API}
\sysname{} provides Python 2 and 3 modules that can be used to do sampling from within a GNN.
For this, users first define \sysname{} API functions, then call \texttt{doSampling} function to do \transit-parallel sampling, and finally call \texttt{getFinalSamples} to obtain samples in a \texttt{numpy.ndarray}.

\subsection{Advantages of \sysname{}'s API}
\addition{
Expressing a graph sampling application using \sysname{}'s API provides several advantages.
It describes sampling operations in a fine-grained manner, which enables using the execution hierarchy of the GPU more efficiently with both sample and transit parallelism.
\sysname{}'s API is general purpose and supports different kinds of graph sampling applications, while existing systems either running specific graph sampling applications~\cite{graphsage,fastgcn,ladies,clustergcn,graphsaint,layersampling} or only support expressing specific kinds of graph sampling applications, as for example random walks for KnightKing~\cite{knightking}.
Furthermore, \sysname{}'s API requires applications to explicitly indicate which vertices in each sample are \transit vertices.
This distinction is critical to enable \transit parallelism.
Finally, the API requires applications to explicitly state the number of vertices that must be sampled at each step.
This information is used by \sysname{} to effectively load balance computation across the execution hierarchy of the GPU.}

\section{Alternative Graph Processing Systems}
\label{sec:alternatives}
Abstractions provided by existing GPU based graph processing systems to implement graph algorithms can be divided into two types:
(i) Message Passing and (ii) Frontier Centric.
We now study the \addition{implementation} of graph sampling using both abstractions and discuss why \addition{these implementations} provides sub-optimal performance.

\spara{Message-passing Abstraction} is provided by several graph computation frameworks~\cite{pregel,tigr,subway,map-graph,cusha}, where each vertex is associated with a local state and a vertex can send messages to its neighbors.
Upon receiving messages, vertices update their state and can send new messages.
Graph computations written in this abstraction advances the computation by exchanging messages between vertices at each step.

A \transit-parallel approach for graph sampling implemented using message passing works in the following way. 
First, in each step for each sample associated with a \transit, neighbors of the \transit are sampled.
\addition{Then, the \texttt{stepTransits} function is called to retrieve \transit for next step and the associated samples are send to the \transit in the form of messages.}
Each \transit vertex is associated with only one thread, which processes all its samples sequentially.

\spara{Frontier-centric Abstraction} is provided by Gunrock~\cite{gunrock}, which exploits the property that after any step of a graph computation, a set of \emph{frontier} vertices are produced for the next step of the computation. 
The \textsc{Advance} operator in this abstraction defines the computation and generates a new frontier by assigning one thread to each neighbor of each vertex in the input frontier.

A \transit-parallel approach for graph sampling can be implemented in this abstraction as follows way.
The \textsc{Advance} operator contains the user-defined sampling criteria, which is invoked on each neighbor of the transit vertex. 
This operator will decide whether the neighbor should be added to the sample. 
\addition{In that case, \texttt{stepTransits} is called to retrieve the \transit vertex for the sample and the \transit vertex is added to the new frontier.}
Each thread for a neighbor must make this decision for all the associated samples, which are processed sequentially.

\spara{Limitations over \sysname} Graph processing systems providing above abstractions suffers from two fundamental issues.
First, these systems only consider one degree of parallelism, i.e., all \transit vertices can be processed in parallel but samples for each \transit are processed sequentially.
This is because these systems are designed for traditional graph computations, such as Breadth First Search, Connected Components, etc., which only has one degree of parallelism.
Second, these systems balances the load based on the number of neighbors for each vertex because all neighbors of each vertex are visited in a traditional graph processing application.
But, in a graph sampling number of neighbors to sample at a step can be significantly less than the neighbors of a \transit vertex. 
Hence, \sysname takes a different approach from existing system to solve these issues.
Section~\ref{sec:experiments:alternatives} shows that \sysname performs better than these systems.

\section{Evaluation}
\label{sec:eval}
We implemented \sysname in NVIDIA CUDA 11.2.

\spara{Benchmarks} We use the graph sampling applications mentioned in Section~\ref{sec:api} as benchmarks for our evaluation.
We set applications' parameters as follows.
For \textit{PPR} the termination probability is set to 1/100, i.e., mean length is 100.
For all other random walks, we set the walk length to 100. 
For \textit{node2vec} we set $p$ to 2.0 and $q$ to 0.5.
For these random walks, initially there is one vertex per sample.
\addition{For \textit{MultiDimensional Random Walk} (MultiRW), we set 100 root vertices per sample.
We use GraphSAGE~\cite{graphsage}'s hyperparameters for \emph{k-hop Neighborhood Sampling}, i.e., $k = 2$, $m_1 = 25$, and $m_2 = 10$.
For \textit{Layer Sampling} we set final sample size to 2000 and step size for all steps to 1000.
For \textit{FastGCN}, \textit{LADIES}, and \textit{MVS} Sampling batch size and step size are set to 64. 
For \textit{ClusterGCN Sampling} we randomly assigned vertices in clusters and each sample contains 20 clusters.}

\spara{Datasets} Table~\ref{tab:graphs} lists the details of real world graphs used in our evaluation obtained from
Stanford Network Analysis Project~\cite{snapnets}.
We generate a weighted version of these graphs by assigning weights to each edge randomly from [1, 5).
\begin{table}[t]
  \footnotesize
  \begin{tabularx}{\columnwidth}{llccc}
      \toprule
      Name & Abrv &\# of Nodes & \# of Edges & \addition{Avg Degree}\\ \hline
      Protein-Protein & PPI & 50K & 1.4M     & \addition{28.0}\\ 
      Interactions\\
      com-Orkut& Orkut &3M&117M              & \addition{39.0}\\
      cit-Patents& Patents &3.77M & 16.5M    & \addition{4.37}\\
      soc-LiveJournal1 & LiveJ & 4.8M & 68.9M & \addition{14.3}\\
      com-Friendster& FriendS & 65.6M & 1.8B   & \addition{27.4}\\
      \bottomrule
  \end{tabularx}
  \caption{Graph used in our evaluation.\vspace{-2em}}
  \label{tab:graphs}
\end{table}

\spara{Experimental setup} We perform experiments on a system containing two 16-core Intel Xeon(R) Silver 4216 CPU, 128 GB RAM, and an NVIDIA Tesla V100 GPU with 16GB memory running Ubuntu 18.04.
We report the average time of 10 executions.
We report the execution time spent on the GPU, which includes the time spent in sampling and creating the scheduling index.
Since transferring graphs to the GPU that fits inside the GPU memory takes only few milliseconds (less than 5\% of total execution time), we do not consider these times in the total execution time unless specified otherwise.

\subsection{Execution Time Breakdown}
The execution time of an application in \sysname{} consists of the time spent in sampling and creating the \textit{scheduling index}.
\sysname{} builds the scheduling index by sorting the samples based on the neighbors in each sample as keys
and then dividing the \transit vertices into three sets based on the number of samples for each \transit, using parallel scan.
Figure~\ref{fig:invert-time} shows the time spent in both phases as a fraction of the total execution time.
The time spent in building scheduling index ranges \addition{from 5\% of the total time in ClusterGCN} for sampling LiveJ graph to 40.4\% of the total time in DeepWalk for sampling Orkut graph.
Random walks spend a higher fraction of time building the scheduling index.
This is because they sample only a single vertex per step, leading to fewer common samples and less work per \transit than other applications.
\sysname{} uses parallel radix sort and parallel scan of NVIDIA CUB~\cite{nvidia-cub}  to create the scheduling index efficiently.
With more efficient implementations of these algorithms~\cite{hybrid-radix-sort} available for GPUs, we expect this time to decrease significantly in future.
\usepgfplotslibrary{units} 

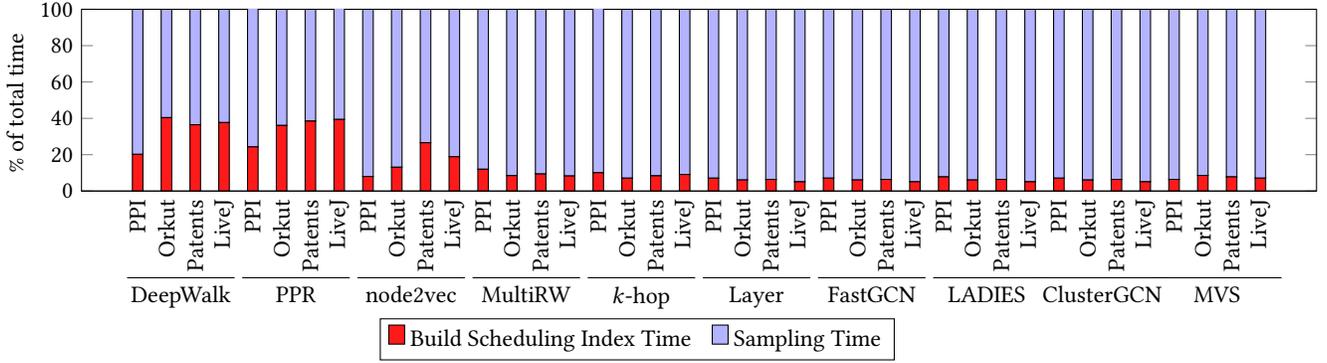
\begin{figure*}[h]
  \small
\pgfplotstableread{
  X   Application Dataset    trans_time   algo_time
  1   DeepWalk    PPI        20.2         79.8
  2   DeepWalk    Orkut      40.4         59.6
  3   DeepWalk    Patents    36.5         63.5
  4   DeepWalk    LJ1        37.7         62.3
  5   PPR         PPI        24.3         76.7
  6   PPR         Orkut      36.1         63.9
  7   PPR         Patents    38.5         61.5
  8   PPR         LJ1        39.4         61.6
  9   node2vec    PPI        8.0          92.0
  10  node2vec    Orkut      13.1         86.9
  11  node2vec    Patents    26.6         73.4
  12  node2vec    LJ1        18.9         81.1
  13  MultiRW    PPI        12.0          88.0
  14  MultiRW    Orkut      8.5           91.5
  15  MultiRW    Patents    9.4           90.6
  16  MultiRW    LJ1        8.3           91.7
  17  khop    PPI            10.1         90.9
  18  khop    Orkut          7.1          92.9
  19  khop    Patents        8.4          91.6
  20  khop    LJ1            9.1          90.9
  21  Layer    PPI            7.1         92.9
  22  Layer    Orkut          6.1          93.9
  23  Layer    Patents        6.4          93.6
  24  Layer    LJ1            5.1          94.9
  25  FastGCN    PPI            7.1         92.9
  26  FastGCN    Orkut          6.1          93.9
  27  FastGCN    Patents        6.4          93.6
  28  FastGCN    LJ1            5.1          94.9
  29  LADIES    PPI            7.8         92.2
  30  LADIES    Orkut          6.1          93.9
  31  LADIES    Patents        6.4          93.6
  32  LADIES    LJ1            5.1          94.9
  33  ClusterGCN    PPI        7.1         92.9
  34  ClusterGCN    Orkut      6.1          93.9
  35  ClusterGCN    Patents    6.4          93.6
  36  ClusterGCN    LJ1        5.1          94.9
  37  MVS    PPI               6.4         93.6
  38  MVS    Orkut             8.6         91.4
  39  MVS    Patents           7.8         92.2
  40  MVS    LJ1               7.1         92.9 
}\datatable

\begin{tikzpicture}
\begin{axis}[
  width=18.0cm, height=4cm, compat=newest,
  ybar stacked,
    ylabel=label,
    xtick=data,
    xticklabels={PPI, Orkut, Patents, LiveJ,PPI, Orkut, Patents, LiveJ,PPI, Orkut, Patents, LiveJ,PPI, Orkut, Patents, LiveJ,PPI, Orkut, Patents, LiveJ
    ,PPI, Orkut, Patents, LiveJ,PPI, Orkut, Patents, LiveJ,PPI, Orkut, Patents, LiveJ,PPI, Orkut, Patents, LiveJ,PPI, Orkut, Patents, LiveJ},
    x tick label style={rotate=90},
    enlarge x limits = 0.05,
    ylabel style={align=center},
    ylabel = \% of total time,
    ymin=0,
    ymax=100,
    bar width=4.0pt,
    legend style={
      font=\small,
      cells={anchor=west},
      legend columns=5,
      at={(0.45,-0.7)},
      anchor=north,
      /tikz/every even column/.append style={column sep=0.2cm}
    },
    legend cell align=left,
    after end axis/.code={\draw ({rel axis cs:0,0}|-{axis cs:0,0}) -- ({rel axis cs:1,0}|-{axis cs:0,0});
    },
    draw group line={[index]1}{DeepWalk}{DeepWalk}{-8.5ex}{4pt},
    draw group line={[index]1}{PPR}{PPR}{-8.5ex}{4pt},
    draw group line={[index]1}{node2vec}{node2vec}{-8.5ex}{4pt},
    draw group line={[index]1}{MultiRW}{\addition{MultiRW}}{-8.5ex}{4pt},
    draw group line={[index]1}{khop}{$k$-hop}{-8.5ex}{4pt},
    draw group line={[index]1}{Layer}{Layer}{-8.5ex}{4pt},
    draw group line={[index]1}{FastGCN}{\addition{FastGCN}}{-8.5ex}{4pt},
    draw group line={[index]1}{LADIES}{\addition{LADIES}}{-8.5ex}{4pt},
    draw group line={[index]1}{ClusterGCN}{\addition{ClusterGCN}}{-8.5ex}{8pt},
    draw group line={[index]1}{MVS}{\addition{MVS}}{-8.5ex}{8pt}
]

\addplot [black,fill=red!90!white] table[x=X,y=trans_time] {\datatable};
\addlegendentry{Build Scheduling Index Time}
\addplot [black,fill=blue!30!white] table[x=X,y=algo_time] {\datatable};
\addlegendentry{Sampling Time}

\end{axis}
\end{tikzpicture}
\caption{Percentage of time spent in Sampling the graph and building Scheduling Index to the total time in \sysname{}\label{fig:invert-time}}
\end{figure*}

\subsection{Graph Sampling Performance}
We compare \sysname with the following systems.

\spara{SP}
\sysname is the first system for graph sampling on GPUs.
Since we cannot compare it with other systems, we implemented an optimized sample-parallel graph sampling system based on the \sysname API, which
we refer to as \emph{SP}.
We implemented all the optimizations of \sysname that could be adapted to a sample-parallel system, such as load balancing, scheduling, and the fine-grained parallelism discussed in Section~\ref{sec:sample-parallel}.
\addition{The purpose of SP is to evaluate the benefits of transit-parallelism in isolation, without considering all the other optimizations enabled by the new API.}

Since TensorFlow based reference implementation of layer sampling~\cite{layersampling} does not support the datasets used in evaluation and the implementation follows sample-parallel paradigm, we use SP as a baseline in layer sampling.

\spara{TP}
To show the performance improvement due to \sysname{}'s load balancing and scheduling optimizations described in Section~\ref{sec:system}, we compare against vanilla \transit-parallel approach (see Section~\ref{sec:paradigm:transit-parallel}), which assigns each \transit and sample pair to $m_i$ consecutive threads.
We refer to this implementation as \emph{TP}.

\spara{KnightKing} KnightKing~\cite{knightking} is a state of the art system for doing random walks using CPUs.
It uses rejection sampling as a technique to select new vertices of a random walk and supports sampling using distributed systems.
Its API restricts expressing only random walks, hence, we use the system as a baseline only for random walks.

\spara{Existing GNN Samplers}\addition{We compare against the samplers of existing GNNs. 
These samplers are written for TensorFlow or numpy and are designed to run only on multi-core CPUs, not GPUs.
This is because sampling is an irregular computation that is more easily implemented on CPUs. 
For $k$-hop neighborhood, we compare against GraphSAGE's sampler~\cite{graphsage}.
For MultiRW, we compare against GraphSAINT's sampler~\cite{graphsaint}.
For sampling algorithms in FastGCN~\cite{fastgcn}, ClusterGCN~\cite{clustergcn}, MVS~\cite{mvs}, and LADIES~\cite{ladies}, we compare against samplers in their reference implementations.
}

\begin{figure*}[t]
  \small
  \begin{subfigure}{0.41\linewidth}
\pgfplotstableread{
  X   Application Dataset    CPURefImpl
  1   DeepWalk    PPI        36.7         
  2   DeepWalk    Orkut      26.1         
  3   DeepWalk    Patents    29.1         
  4   DeepWalk    LJ1        38.9         
  5   PPR         PPI        28.2         
  6   PPR         Orkut      26.2         
  7   PPR         Patents    27.0        
  8   PPR         LJ1        31.1         
  9   node2vec    PPI        7.42         
  10  node2vec    Orkut      29.7         
  11  node2vec    Patents    46.7         
  12  node2vec    LJ1       50              
}\datatable
\begin{tikzpicture}
\begin{axis}[
  width=8.0cm, height=4.3cm, compat=newest,
    ylabel=label,
    xtick=data,
    xticklabels={PPI, Orkut, Patents, LiveJ,PPI, Orkut, Patents, LiveJ,PPI, Orkut, Patents, LiveJ,PPI, Orkut, Patents, LiveJ},
    x tick label style={rotate=90},
    enlarge x limits = 0.05,
    ylabel style={align=center},
    ylabel = {Speedup},
    ybar=0pt,
    ymin=0,
    ymax=60,
    bar width=4.0pt,
    legend style={
      font=\small,
      cells={anchor=west},
      legend columns=5,
      at={(0.45,-0.75)},
      anchor=north,
      /tikz/every even column/.append style={column sep=0.2cm}
    },
    legend cell align=left,
    after end axis/.code={\draw ({rel axis cs:0,0}|-{axis cs:0,0}) -- ({rel axis cs:1,0}|-{axis cs:0,0});
        \path [anchor=base east, yshift=0.5ex]
            ;
    },
    nodes near coords,
    every node near coord/.append style={font=\scriptsize, rotate=90, anchor=west},
    nodes near coords align={vertical},
    draw group line={[index]1}{DeepWalk}{DeepWalk}{-10ex}{4pt},
    draw group line={[index]1}{PPR}{PPR}{-10ex}{4pt},
    draw group line={[index]1}{node2vec}{node2vec}{-10ex}{4pt},
    draw group line={[index]1}{MultiRW}{MultiRW}{-10ex}{4pt},
]

\addplot [red!40!black,fill=red!90!white] table[x=X,y=CPURefImpl] {\datatable};

\end{axis}
\end{tikzpicture}
\caption{Speedup of on random walks over KnightKing.}
\label{fig:speed-rand-walks}
\end{subfigure}
\hspace{1em}
\begin{subfigure}{0.56\linewidth}
  \pgfplotstableread{
    X   Application Dataset    CPURefImpl
    1  khop    PPI            2.000      
    2  khop    Orkut          1.200        
    3  khop    Patents        1.570         
    4  khop    LJ1            1.475                   
    5  FastGCN    PPI          0.187         
    6  FastGCN    Orkut        1.532         
    7  FastGCN    Patents      1.223         
    8  FastGCN    LJ1          2.010         
    9  LADIES    PPI           0.500         
    10  LADIES    Orkut         1.200       
    11  LADIES    Patents       2.300          
    12  LADIES    LJ1           1.420         
    13  ClusterGCN    PPI      0.500       
    14  ClusterGCN    Orkut     1.520          
    15  ClusterGCN    Patents   1.405          
    16  ClusterGCN    LJ1       1.304     
    17  MVS    PPI              0.800   
    18  MVS    Orkut            1.023
    19  MVS    Patents          1.100   
    20  MVS    LJ1              1.200  
    21  MultiRW    PPI          0.020         
    22  MultiRW    Orkut      0.021       
    23  MultiRW    Patents    0.024         
    24  MultiRW    LJ1        0.020   
  }\datatable
  \begin{tikzpicture}
  \begin{axis}[
    width=10.0cm, height=4.3cm, compat=newest,
      ylabel=label,
      xtick=data,
      xticklabels={PPI, Orkut, Patents, LiveJ
      ,PPI, Orkut, Patents, LiveJ,PPI, Orkut, Patents, LiveJ,PPI, Orkut, Patents, LiveJ,PPI, Orkut, Patents, LiveJ,PPI, Orkut, Patents, LiveJ},
      x tick label style={rotate=90},
      enlarge x limits = 0.05,
      ylabel style={align=center},
      ylabel = {Speedup ($\times$ 10$^3$)},
      ybar=0pt,
      ymin=0,
      ymax=3,
      bar width=4.0pt,
      legend style={
        font=\small,
        cells={anchor=west},
        legend columns=5,
        at={(0.45,-0.75)},
        anchor=north,
        /tikz/every even column/.append style={column sep=0.2cm}
      },
      legend cell align=left,
      after end axis/.code={\draw ({rel axis cs:0,0}|-{axis cs:0,0}) -- ({rel axis cs:1,0}|-{axis cs:0,0});
          \path [anchor=base east, yshift=0.5ex]
              ;
      },
      nodes near coords,
      every node near coord/.append style={font=\scriptsize, rotate=90, anchor=west},
      nodes near coords align={vertical},
      draw group line={[index]1}{DeepWalk}{DeepWalk}{-10ex}{4pt},
      draw group line={[index]1}{PPR}{PPR}{-10ex}{4pt},
      draw group line={[index]1}{node2vec}{node2vec}{-10ex}{4pt},
      draw group line={[index]1}{MultiRW}{\addition{MultiRW}}{-10ex}{4pt},
      draw group line={[index]1}{khop}{$k$-hop}{-10ex}{4pt},
      draw group line={[index]1}{Layer}{Layer}{-10ex}{4pt},
      draw group line={[index]1}{FastGCN}{\addition{FastGCN}}{-10ex}{4pt},
      draw group line={[index]1}{LADIES}{\addition{LADIES}}{-10ex}{4pt},
      draw group line={[index]1}{ClusterGCN}{\addition{ClusterGCN}}{-10ex}{4pt},
      draw group line={[index]1}{MVS}{\addition{MVS}}{-10ex}{4pt},
  ]
  
  \addplot [red!40!black,fill=red!90!white] table[x=X,y=CPURefImpl] {\datatable};
  
  \end{axis}
  \end{tikzpicture}
  \caption{\addition{Speedup on sampling applications over their GNN implementation ($\times\, 10^3$).}}
  \label{fig:speedup-gnn-apps}
  \end{subfigure}

  \begin{subfigure}{\linewidth}
      \pgfplotstableread{
        X   Application Dataset    SP        TP
        1   DeepWalk    PPI        2.1         2.18
        2   DeepWalk    Orkut      1.76         2.03
        3   DeepWalk    Patents    2.10         2.14
        4   DeepWalk    LJ1        2.27         2.15
        5   PPR         PPI        2.02         1.98
        6   PPR         Orkut      1.78         1.71
        7   PPR         Patents    1.65         1.71
        8   PPR         LJ1        2.38         2.34
        9   node2vec    PPI        2.13         2.08
        10  node2vec    Orkut      1.09         1.02
        11  node2vec    Patents    1.21         1.23
        12  node2vec    LJ1        1.40         1.36
        13  MultiRW    PPI         1.1          1.25
        14  MultiRW    Orkut       1.26         1.30
        15  MultiRW    Patents     1.24         1.32
        16  MultiRW    LJ1         1.31         1.37
        17  khop       PPI         4.5         2.5
        18  khop    Orkut          3.55         1.67
        19  khop    Patents        2.25         1.27
        20 khop    LJ1            3.2          2.3
        21  Layer    PPI           4.36         2.26
        22 Layer    Orkut         5.10         2.55
        23 Layer    Patents       3.88         1.94
        24  Layer    LJ1           2.33         2.00
        25  FastGCN    PPI         2.5          1.7
        26  FastGCN    Orkut       2.3          1.6
        27  FastGCN    Patents     2.1          1.4
        28  FastGCN    LJ1         2.4          1.5
        29  LADIES    PPI          2.5          1.6
        30  LADIES    Orkut        2.3          1.8
        31  LADIES    Patents      2.1          1.5
        32  LADIES    LJ1          2.4          1.5
        33  ClusterGCN    PPI      2.5          1.4
        34  ClusterGCN    Orkut    2.4          1.9
        35  ClusterGCN    Patents  2.1          1.6
        36  ClusterGCN    LJ1      2.7          1.8    
        37  MVS    PPI             2.1          1.9
        38  MVS    Orkut           2.9          2.0
        39  MVS    Patents         2.0          1.7
        40  MVS    LJ1             2.1          1.7 
      }\datatable
      \begin{tikzpicture}
      \begin{axis}[
        width=18.0cm, height=4.7cm, compat=newest,
          ylabel=label,
          xtick=data,
          xticklabels={PPI, Orkut, Patents, LiveJ,PPI, Orkut, Patents, LiveJ,PPI, Orkut, Patents, LiveJ,PPI, Orkut, Patents, LiveJ,PPI, Orkut, Patents, LiveJ
          ,PPI, Orkut, Patents, LiveJ,PPI, Orkut, Patents, LiveJ,PPI, Orkut, Patents, LiveJ,PPI, Orkut, Patents, LiveJ,PPI, Orkut, Patents, LiveJ},
          x tick label style={rotate=90},
          enlarge x limits = 0.025,
          ylabel style={align=center},
          ylabel = {Speedup},
          ybar=2pt,
          x=0.42cm,
          ymin=0,
          ymax=6,
          ytick={0,1,2,3,4,5,6},
          bar width=4.0pt,
          legend style={
            font=\small,
            cells={anchor=west},
            legend columns=5,
            at={(0.9,0.8)},
            anchor=south,
            /tikz/every even column/.append style={column sep=0.2cm}
          },
          legend cell align=left,
          after end axis/.code={\draw ({rel axis cs:0,0}|-{axis cs:0,0}) -- ({rel axis cs:1,0}|-{axis cs:0,0});
              \path [anchor=base east, yshift=0.5ex]
                  ;
          },
          nodes near coords,
          every node near coord/.append style={font=\scriptsize, rotate=90, anchor=west},
          nodes near coords align={vertical},
          draw group line={[index]1}{DeepWalk}{DeepWalk}{-10ex}{4pt},
          draw group line={[index]1}{PPR}{PPR}{-10ex}{4pt},
          draw group line={[index]1}{node2vec}{node2vec}{-10ex}{4pt},
          draw group line={[index]1}{MultiRW}{MultiRW}{-10ex}{4pt},
          draw group line={[index]1}{khop}{$k$-hop}{-10ex}{4pt},
          draw group line={[index]1}{Layer}{Layer}{-10ex}{4pt},
          draw group line={[index]1}{FastGCN}{\addition{FastGCN}}{-10ex}{4pt},
          draw group line={[index]1}{LADIES}{\addition{LADIES}}{-10ex}{4pt},
          draw group line={[index]1}{ClusterGCN}{\addition{ClusterGCN}}{-10ex}{4pt},
          draw group line={[index]1}{MVS}{\addition{MVS}}{-10ex}{4pt},
      ]
      
      \addplot [red!40!black,fill=red!90!white] table[x=X,y=SP] {\datatable};
      \addlegendentry{SP}
      \addplot [red!40!black,fill=blue!90!white] table[x=X,y=TP] {\datatable};
      \addlegendentry{TP}
      
      \end{axis}
      \end{tikzpicture}
      \caption{\addition{Speedup on graph sampling applications over SP and TP.}}
      \label{fig:speedup-sp-tp}
      \end{subfigure}
  \caption{Speedup of \sysname{} on graph sampling applications and real world graphs over baselines.}
\end{figure*}
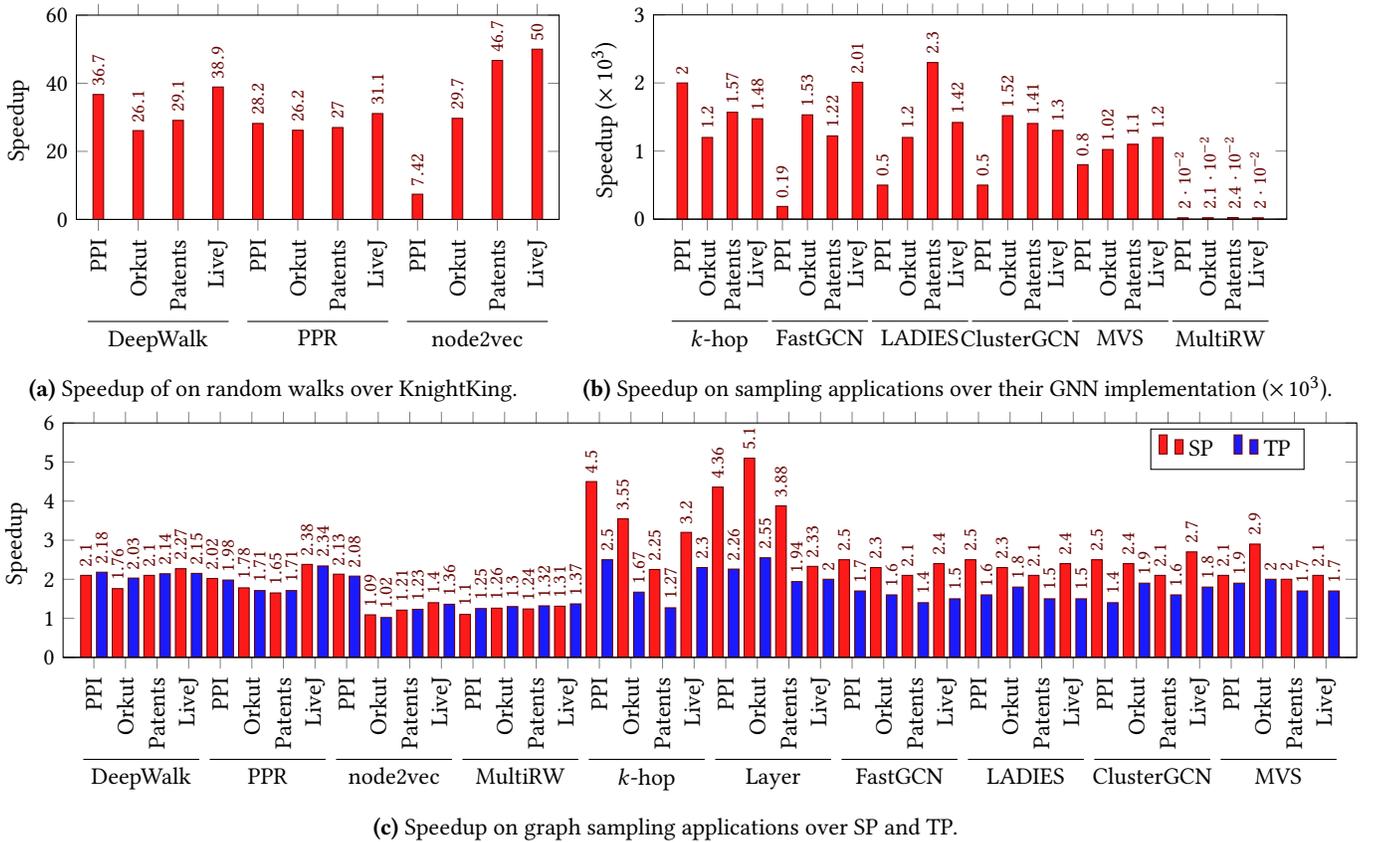

\pgfplotsset{height=4cm, compat=newest} 
\usepgfplotslibrary{units} 

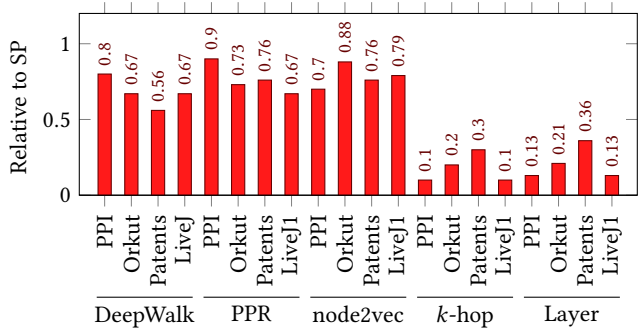
\begin{figure}[t]
  \small
\pgfplotstableread{
  X   Application Dataset    gld_trans      warp_exec_eff    l1_hit_rate
  1   DeepWalk    PPI        0.8        1.0              1.89
  2   DeepWalk    Orkut      0.67           1.35             2.21
  3   DeepWalk    Patents    0.56           0.98             4.30
  4   DeepWalk    LJ1        0.67           1.12             2.21
  5   PPR         PPI        0.9            1.1              1.60
  6   PPR         Orkut      0.73           1.40             2.40
  7   PPR         Patents    0.76           1.10             3.89
  8   PPR         LJ1        0.67           1.10             2.34
  9   node2vec    PPI        0.7           1.48             0.95
  10   node2vec   Orkut      0.88           1.25             1.0
  11   node2vec   Patents    0.76           1.50             5.4
  12   node2vec   LJ1        0.79           1.72             1.1
  13   khop       PPI        0.1            1.5              0.95
  14   khop       Orkut      0.2            1.7              1.0
  15   khop       Patents    0.3            1.50             5.4
  16   khop       LJ1        0.1            4.72             1.1
  17   Layer       PPI        0.13          1.9              0.95
  18   Layer       Orkut      0.21          1.4              2.0
  19   Layer       Patents    0.36          1.8             3.4
  20   Layer       LJ1        0.13          2.6              2.1
}\datatable

\begin{tikzpicture}
\begin{axis}[width=9cm,
    ylabel=label,
    xtick=data,
    xticklabels={PPI, Orkut, Patents, LiveJ,PPI, Orkut, Patents, LiveJ1,PPI, Orkut, Patents, LiveJ1,PPI, Orkut, Patents, LiveJ1,PPI, Orkut, Patents, LiveJ1},
    x tick label style={rotate=90},
    enlarge x limits = 0.05,
    ylabel style={align=center},
    ylabel = Relative to SP,
    ybar=0pt,
    ymin=0,
    ymax=1.2,
    bar width=5.0pt,
    legend style={
      font=\small,
      cells={anchor=west},
      legend columns=5,
      at={(0.45,-0.63)},
      anchor=north,
      /tikz/every even column/.append style={column sep=0.2cm}
    },
    legend cell align=left,
    after end axis/.code={\draw ({rel axis cs:0,0}|-{axis cs:0,0}) -- ({rel axis cs:1,0}|-{axis cs:0,0});
        \path [anchor=base east, yshift=0.5ex]
            ;
    },
    nodes near coords,
    every node near coord/.append style={font=\scriptsize, rotate=90, anchor=west},
    nodes near coords align={vertical},
    draw group line={[index]1}{DeepWalk}{DeepWalk}{-10ex}{3pt},
    draw group line={[index]1}{PPR}{PPR}{-10ex}{3pt},
    draw group line={[index]1}{node2vec}{node2vec}{-10ex}{3pt},
    draw group line={[index]1}{khop}{$k$-hop}{-10ex}{3pt},
    draw group line={[index]1}{Layer}{Layer}{-10ex}{3pt},
]

\addplot [red!40!black,fill=red!90!white] table[x=X,y=gld_trans] {\datatable};

\end{axis}
\end{tikzpicture}
\caption{Value of different L2 Cache Load Transactions metric for several applications in \sysname{} relative to SP.\label{fig:perf-metrics}}
\end{figure}

\spara{Performance Results}
\sysname{} provides an order of magnitude speedup over KnightKing (Figure~\ref{fig:speed-rand-walks}) for all random walk applications, with speedups ranging from 26.1$\times$ to 50$\times$.
\addition{\sysname{} provides an order of magnitude speedup over the implementations of existing GNNs (Figure~\ref{fig:speedup-gnn-apps}).}
These large speedups are possible due to the massive parallelism and memory access latency hiding capabilities of the GPU.
Furthermore, SP is significantly faster than all baselines.

\sysname{} provides significant speedups over SP on all graph sampling applications, with speedups ranging from 1.09$\times$ to 6$\times$.
The speedup depends significantly on the application.
For example, \sysname{} obtains more speedup in DeepWalk and PPR than in node2vec
because in node2vec at each step, for an edge from current transit vertex $v$ to a vertex $u$, the algorithm might do a search over the edges of the previous transit vertex
$t$ to check if $u$ is a neighbor of $t$,
leading to memory accesses and warp divergence.
Nevertheless, \sysname{} still obtains speedup due to its \transit-parallel paradigm.
\sysname{} achieves speedup over SP in all applications because \sysname{} uses three levels of parallelism
while SP can use only two levels of parallelism.
\addition{Moreover, with FastGCN and LADIES, \sysname{} is faster because it speeds up the computation of the combined neighborhood.}

\sysname{} significantly improves performance over TP due to better load balancing and scheduling.
TP is competitive to SP in random walks even though significant time is spent in map inversion because TP caches the neighbors in shared memory.
TP outperforms SP in other applications because caching neighbors in shared memory decreases memory access time when sampling many neighbors.

\subsubsection{\sysname{}'s Effectiveness over SP}
To explain \\\sysname{}'s effectiveness over SP, we obtained values of L2 Cache Read Transaction performance metrics using \texttt{nvprof}.
This metric represents the total number of L2 cache load transactions in the entire execution.
Figure~\ref{fig:perf-metrics} shows the value of this metrics for \sysname{} relative to SP.
\sysname{} performs a fraction of the transactions of SP because it performs coalesced reads and caches edges of \transit vertices in shared memory and registers.
ClusterGCN, MVS, FastGCN, and LADIES sampling perform a similar number of loads and stores as $k$-hop and Layer sampling.

\subsubsection{\sysname{}'s Efficiency}
We present absolute values of two performance metrics in Table~\ref{tab:sysname-efficiency} obtained using \texttt{nvprof}: \addition{(i) \emph{Global Memory Store Efficiency} to show \sysname{}'s effectiveness to do efficient global stores}, and (ii) \emph{Multiprocessor Activity} to show \sysname{}'s effectiveness in fully utilizing GPU's execution resources.

\addition{
\spara{Global Memory Store Efficiency} is the ratio of extra store transactions over the ideal number of transactions to the ideal number of transactions.
Hence, higher efficiency is better.
\sysname{} performs fully efficient global memory stores because of the sub-warp execution.
Since ClusterGCN, MVS, FastGCN, and LADIES sampling perform number of loads and stores similar to $k$-hop and Layer sampling, we found similar store efficiency for these applications.
}

\begin{table}[t]
  \small
  \begin{tabularx}{\linewidth}{l|cc|ccccc}
      \toprule

      Dataset & \multicolumn{2}{c|}{Store} & \multicolumn{5}{c}{Multiprocessor Activity(\%)} \\
              & \multicolumn{2}{c|}{Efficiency(\%)}&  \\ \hline
              & $k$-hop & Layer & DW& PPR & n2v & $k$-hop & Layer\\ \hline
      PPI     &  98.5    & 98.5   & 67.8 & 69.8 & 70.1 & 100 & 100\\
      Orkut   &  99.5    & 100   & 98.3& 98.0 & 99.3  & 100&100\\
      Patents &  100    &  100   & 90.1  & 99.0 & 99.0 & 100&100\\
      LiveJ   &  100    &  100   & 99.2 & 98.2& 97.6 & 100&100\\
      \bottomrule
  \end{tabularx}
  \caption{\addition{Global Memory Store Efficiency} and Multiprocessor Activity in \sysname{}. (DW is short for DeepWalk and n2v is short for node2vec) \label{tab:sysname-efficiency} \vspace{-2em}}
\end{table}

\spara{Multiprocessor Activity} is the average usage of all SMs over the entire execution of the application.
For PPI, Multiprocessor Activity is low because PPI is a small graph and not enough threads are generated to fully utilize all SMs. For all graphs \sysname{} fully utilizes all SMs.
Hence, \sysname{}'s load balancing strategy balances load across all SMs.
\addition{We found similar results for other sampling applications.}

\subsection{Alternative GPU-Based Abstractions}
\label{sec:experiments:alternatives}
We also compare \sysname{} with two state-of-the-art graph processing frameworks: Gunrock~\cite{gunrock} and Tigr~\cite{tigr}, which follow the frontier-centric and message-passing abstractions, respectively (see Section~\ref{sec:alternatives}).
Figure~\ref{fig:abstractions} reports the speedup of \sysname.
As explained in Section~\ref{sec:alternatives}, low parallelism and poor load balancing due to the mismatch between graph sampling and graph processing abstraction result in speedup.
\addition{We found similar results on other applications.}

\pgfplotsset{width=9.0cm, height=4cm, compat=newest} 
\usepgfplotslibrary{units} 

\begin{figure}[t]
  \small
\pgfplotstableread{
  X   Application Dataset    Gunrock  Tigr
  1   DeepWalk    PPI        4.6      6.5
  2   DeepWalk    Orkut      5.9      6.9
  3   DeepWalk    Patents    5.7      5.5
  4   DeepWalk    LJ1        6.7      8.9
  5   PPR         PPI        8.5      8.6
  6   PPR         Orkut      7.6      7.4
  7   PPR         Patents    6.8      6.3
  8   PPR         LJ1        5.6      6.5
  9   node2vec    PPI        5.1      4.7
  10   node2vec   Orkut      4.1      4.0
  11   node2vec   Patents    6.5      6.9
  12   node2vec   LJ1        5.5      4.0
  13   khop       PPI        4.5      6.0
  14   khop       Orkut      5.7      3.5
  15   khop       Patents    5.4      2.2
  16   khop       LJ1        4.5      6.0
  17   Layer       PPI        4.5      6.0
  18   Layer       Orkut      5.7      3.5
  19   Layer       Patents    5.4      2.2
  20   Layer       LJ1        4.5      6.0
}\datatable
\begin{tikzpicture}
\begin{axis}[
    ylabel=label,
    xtick=data,
    xticklabels={PPI, Orkut, Patents, LiveJ,PPI, Orkut, Patents, LiveJ,PPI, Orkut, Patents, LiveJ,PPI, Orkut, Patents, LiveJ,PPI, Orkut, Patents, LiveJ},
    x tick label style={rotate=90},
    enlarge x limits = 0.05,
    ylabel style={align=center},
    ylabel = {Speedup},
    ybar=0pt,
    ymin=0,
    ytick={0,2,4,6,8,10},
    bar width=4.0pt,
    legend style={
      font=\small,
      cells={anchor=west},
      legend columns=5,
      at={(0.77,0.72)},
      anchor=south,
      /tikz/every even column/.append style={column sep=0.2cm}
    },
    legend cell align=left,
    after end axis/.code={\draw ({rel axis cs:0,0}|-{axis cs:0,0}) -- ({rel axis cs:1,0}|-{axis cs:0,0});
        \path [anchor=base east, yshift=0.5ex]
            ;
    },
    nodes near coords align={vertical},
    draw group line={[index]1}{DeepWalk}{DeepWalk}{-10ex}{4pt},
    draw group line={[index]1}{PPR}{PPR}{-10ex}{4pt},
    draw group line={[index]1}{node2vec}{node2vec}{-10ex}{4pt},
    draw group line={[index]1}{khop}{$k$-hop}{-10ex}{4pt},
    draw group line={[index]1}{Layer}{Layer}{-10ex}{4pt},
]

\addplot [red!40!black,fill=red!90!white] table[x=X,y=Gunrock] {\datatable};
\addlegendentry{Gunrock}
\addplot [green!40!black,fill=green!90!white] table[x=X,y=Tigr] {\datatable};
\addlegendentry{Tigr}

\end{axis}
\end{tikzpicture}
\caption{Speedup of \sysname{} over Tigr and Gunrock on some graph sampling applications and real world graphs.}
\label{fig:abstractions}
\end{figure}
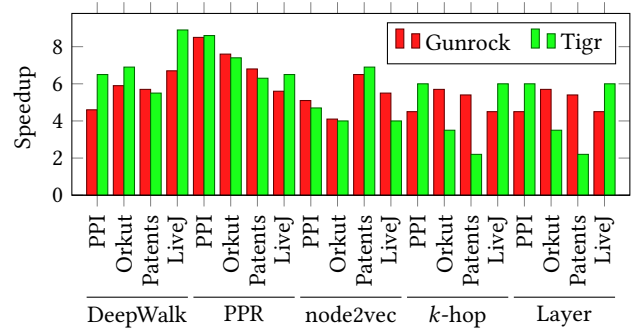

\subsection{Sampling Large Graphs}


We evaluate a simple approach for sampling large graphs.

\sysname{} can sample graphs that do not fit in GPU memory by creating disjoint sub-graphs, such that each of these sub-graphs and its samples be allocated in the GPU memory.
After creating these sub-graphs at each computation step, \sysname{} performs sampling for each sample by transferring all sub-graphs containing the \transit vertices of each sample to the GPU.
In this experiment, we consider the time taken to transfer graph from CPU to GPU.

We evaluate this approach by executing $k$-hop and random walks on the FriendS graph, which does not fit in the GPU memory.
For $k$-hop neighborhood and layer sampling, \sysname is the only system in our experiments that can sample a graph of that size.
\sysname{} gives a throughput of 3.3$\times$10$^{6}$ samples per second on $k$-hop and 2$\times$10$^{6}$ on layer sampling. Both applications are computation bound and not memory transfer bound.
For random walks, KnightKing is the only baseline that can perform the sampling because it is CPU based.
\sysname{} performs worse than KnightKing for random walks where the computation load is low: it provides about $1/2$ of the throughput with DeepWalk and PPR.
However, in node2vec where the computation time is larger, \sysname{} gives 1.50$\times$ speedup over KnightKing.

In summary, \sysname{} is able to sample graphs that do not fit in GPU memory, and can outperform state-of-the-art systems when the graph sampling application performs significant amount of computation.
We plan to improve the support for large graphs in \sysname{} as future work.

\subsection{Sampling on Multiple GPUs}
We used \sysname to perform sampling on four NVIDIA Tesla V100 GPUs.
Figure~\ref{fig:multi-gpu-speedup} presents the speedup of sampling using four GPUs over single GPU.
Multi-GPU sampling achieves significant speedup over single GPU on several applications.
Random walks achieves significant speedup in all graphs except PPI because PPI is a small graph.
On the other hand, $k$-hop neighbors achieves almost full scaling even in small graph like PPI because it increases the number of \transit vertices exponentially at each step.
In summary, \sysname is able to utilize multiple GPUs efficiently.

\pgfplotsset{height=3.5cm, compat=newest} 
\usepgfplotslibrary{units} 

\begin{figure}[t]
  \small
\pgfplotstableread{
  X   Application Dataset    speedup
  1   DeepWalk    PPI        1.5
  2   DeepWalk    Orkut      3.1
  3   DeepWalk    Patents    2.7
  4   DeepWalk    LJ1        2.9
  5   PPR         PPI        1.5
  6   PPR         Orkut      3.2
  7   PPR         Patents    2.4
  8   PPR         LJ1        2.6
  9   node2vec    PPI        1.3
  10   node2vec   Orkut      3.3
  11   node2vec   Patents    3.2
  12   node2vec   LJ1        3.1
  13   khop       PPI        3.1
  14   khop       Orkut      3.6
  15   khop       Patents    3.4
  16   khop       LJ1        3.1
  17   Layer       PPI        3.2
  18   Layer      Orkut      3.7
  19   Layer       Patents    3.5
  20   Layer       LJ1        3.3
}\datatable
\begin{tikzpicture}
\begin{axis}[
    ylabel=label,
    xtick=data,
    xticklabels={PPI, Orkut, Patents, LiveJ,PPI, Orkut, Patents, LiveJ,PPI, Orkut, Patents, LiveJ,PPI, Orkut, Patents, LiveJ,PPI, Orkut, Patents, LiveJ},
    x tick label style={rotate=90},
    enlarge x limits = 0.05,
    ylabel style={align=center},
    ylabel = {Speedup},
    ybar=5pt,
    ymin=0,
    ymax=6,
    bar width=5.0pt,
    legend style={
      font=\small,
      cells={anchor=west},
      legend columns=5,
      at={(0.45,-0.63)},
      anchor=north,
      /tikz/every even column/.append style={column sep=0.2cm}
    },
    legend cell align=left,
    after end axis/.code={\draw ({rel axis cs:0,0}|-{axis cs:0,0}) -- ({rel axis cs:1,0}|-{axis cs:0,0});
        \path [anchor=base east, yshift=0.5ex]
            ;
    },
    nodes near coords,
    every node near coord/.append style={font=\scriptsize, rotate=90, anchor=west},
    nodes near coords align={vertical},
    draw group line={[index]1}{DeepWalk}{DeepWalk}{-10ex}{4pt},
    draw group line={[index]1}{PPR}{PPR}{-10ex}{4pt},
    draw group line={[index]1}{node2vec}{node2vec}{-10ex}{4pt},
    draw group line={[index]1}{khop}{$k$-hop}{-10ex}{4pt},
    draw group line={[index]1}{Layer}{Layer}{-10ex}{4pt},
]

\addplot [red!40!black,fill=red!90!white] table[x=X,y=speedup] {\datatable};

\end{axis}
\end{tikzpicture}
\caption{Speedup of sampling using 4 GPUs over 1 GPU.}
\label{fig:multi-gpu-speedup}
\end{figure}
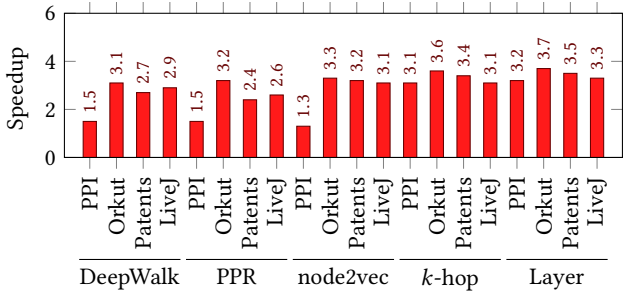

\subsection{End-to-End Integration in GNN Systems}
We performed an end-to-end evaluation of existing GNNs by replacing their sampler with the sampling implementation in \sysname.
Table~\ref{tab:end-to-end} shows the performance improvement of our integration.
The speedup for GraphSAGE is less than the maximum possible improvement in Table~\ref{tab:sampling-overhead} due to a limitation of Tensorflow, which does not allow creating a tensor on the GPU memory.
Therefore, samples are copied to the CPU and then again to the GPU for training.
\addition{For FastGCN and LADIES, the speedup increases with larger graphs because the sampling time depends on the number of vertices in the graph, while the training time per batch remains constant.
Since \sysname{} provides significant speedup over samplers in GraphSAINT and MVS, we believe \sysname{} integration will improve training time.}

\begin{table}[t]
  \begin{tabularx}{\linewidth}{c|ccccc}
    \toprule
            & PPI           & Reddit       & Orkut & Patents      & LiveJ \\
  \hline
  GraphSAGE & 1.30$\times$  & 1.21$\times$ & OOM   & 1.20$\times$ & 1.22$\times$\\
  \addition{FastGCN}   & \addition{1.25$\times$}  & \addition{1.52$\times$} & \addition{4.75$\times$}  & \addition{2.3$\times$}  & \addition{4.31$\times$}\\
  \addition{LADIES}    & \addition{1.07$\times$}  & \addition{1.37$\times$} & \addition{2.27$\times$}  & \addition{2.1$\times$}  & \addition{2.34$\times$}\\
  \addition{ClusterGCN} &\addition{ 1.03$\times$} & \addition{1.20$\times$} & \addition{OOM} & \addition{1.4$\times$} & \addition{1.51$\times$}\\
  \bottomrule
  \end{tabularx}
  \caption{\addition{End-to-end speedups after integrating \sysname in GNNs over vanilla GNNs. \label{tab:end-to-end}} \vspace{-2em}}
\end{table}

%
%
%
%
%
%
%
%


\section{Related Work}
\label{sec:relwork}

We now discuss related work beyond KnightKing, Gunrock, and Tigr, which we discussed in Sections~\ref{sec:alternatives} and~\ref{sec:eval}.

\spara{Message-passing graph processing}
There are several graph processing systems that provide a message-passing abstraction that run on CPUs~\cite{pregel,graphlab,ligra,galois,powergraph,gemini,SympleGraph} and GPUs~\cite{medusa,cusha,map-graph,tigr,subway}.
Our evaluation shows that \sysname outperforms Tigr~\cite{tigr} on  graph sampling tasks (Section~\ref{sec:eval}).
Medusa~\cite{medusa} was the first GPU-based graph processing framework to provide a message passing abstraction.
CuSha \cite{cusha} and MapGraph~\cite{map-graph} provide a Gather And Scatter (GAS) abstraction.
CuSha uses a parallel sliding-window graph representation (``G-Shards'') to avoid irregular memory accesses.
Subway~\cite{subway} splits the large graphs that do not fit in GPU memory into sub-graphs and optimizes memory transfers between CPU and GPU. Shi et al~\cite{graph-GPU-survey} present an extensive review of systems for
graph processing on GPUs.
PowerLyra~\cite{PowerLyra} uses different computations on vertices based on their degree.

\spara{Frontier-centric graph processing}
SIMD-X~\cite{simd-x} extends the frontier abstraction of Gunrock~\cite{gunrock}, but these extensions are not relevant for graph sampling.



\spara{Graph mining}
Graph mining systems follow a subgraph-parallel paradigm that is analogous to sample-parallelism~\cite{nscale,gminer,rstream,asap,fractal,automine,pangolin,arabesque}.
However, even the sample-parallel sampling algorithm of Section~\ref{sec:paradigm} introduces optimizations that are specific to the graph sampling abstraction of Section~\ref{sec:graph-sampling-apps} and do not generalize to graph mining problems.
1)~In graph sampling the number of samples is fixed, whereas graph mining problem may involve exploring an exponential number of subgraphs. 2)~sampling adds a constant number of new vertices to each sample at each step.
This makes it possible to associate new vertices to threads at scheduling time, \emph{before} visiting the graph. 3)~Sampling has a notion of \emph{transit} vertices.
\sysname leverages all these features.

\section{Conclusion}
We show that efficient graph sampling on GPUs is non-trivial.
Existing sampling and graph processing systems do not provide the right abstractions to efficiently support several graph sampling algorithms on GPUs. 
We introduce \emph{\transit-parallel} sampling, a new paradigm for graph sampling that is amenable to an efficient GPU implementation. 
We present \sysname, a system that implements \transit-parallel sampling for GPUs to provide regular memory access and computation, and a high-level API to express several graph sampling applications. 
We show that \sysname{} is significantly faster than the existing systems on several applications.

\paragraph{Acknowledgements}

This work was partially supported by the National
Science Foundation under grant CCF-2102288, a
Facebook Systems for Machine Learning Award, and 
an AWS Cloud Credit for Research grant.

\bibliographystyle{ACM-Reference-Format}
\bibliography{main}

\end{document}